\documentclass[a4paper,12pt]{article}
\usepackage{latexsym}
\makeatletter

\newcommand{\p}[1]{(\ref{#1})}

\def\a{\alpha}

\def\PRL #1 #2 #3{{\em Phys. Rev. Lett. \/} {\bf#1} (#2) #3}
\def\NPB #1 #2 #3{{\em Nucl. Phys. \/} {\bf B#1} (#2) #3}
\def\NPBFS #1 #2 #3 #4{{\em Nucl. Phys. \/} {\bf B#2} [FS#1] (#3) #4}
\def\CMP #1 #2 #3{{\em Commun. Math. Phys. \/} {\bf #1} (#2) #3}
\def\PRD #1 #2 #3{{\em Phys. Rev. \/} {\bf D#1} (#2) #3}
\def\PLA #1 #2 #3{{\em Phys. Lett. \/} {\bf #1A} (#2) #3}
\def\PLB #1 #2 #3{{\em Phys. Lett. \/} {\bf B#1} (#2) #3}
\def\JMP #1 #2 #3{{\em J. Math. Phys. \/} {\bf #1} (#2) #3}
\def\PTP #1 #2 #3{{\em Prog. Theor. Phys. \/} {\bf #1} (#2) #3}
\def\SPTP #1 #2 #3{{\em Suppl. Prog. Theor. Phys. \/} {\bf #1} (#2) #3}
\def\AoP #1 #2 #3{{\em Ann. of Phys. \/} {\bf #1} (#2) #3}
\def\PNAS #1 #2 #3{{\em Proc. Natl. Acad. Sci. USA} {\bf #1} (#2) #3}
\def\RMP #1 #2 #3{{\em Rev. Mod. Phys. \/} {\bf #1} (#2) #3}
\def\PR #1 #2 #3{{\em Phys. Reports \/} {\bf #1} (#2) #3}
\def\AoM #1 #2 #3{{\em Ann. of Math. \/} {\bf #1} (#2) #3}
\def\UMN #1 #2 #3{{\em Usp. Mat. Nauk \/} {\bf #1} (#2) #3}
\def\FAP #1 #2 #3{{\em Funkt. Anal. Prilozheniya \/} {\bf #1} (#2) #3}
\def\FAaIA #1 #2 #3{{\em Functional Analysis and Its Application \/} {\bf
#1} (#2) #3}
\def\BAMS #1 #2 #3{{\em Bull. Am. Math. Soc. \/} {\bf #1} (#2)
#3} \def\TAMS #1 #2 #3{{\em Trans. Am. Math. Soc. \/} {\bf #1}
(#2) #3}
\def\InvM #1 #2 #3{{\em Invent. Math. \/} {\bf #1} (#2) #3}
\def\LMP #1 #2 #3{{\em Letters in Math. Phys. \/} {\bf #1} (#2) #3}
\def\IJMPA #1 #2 #3{{\em Int. J. Mod. Phys. \/} {\bf A#1} (#2) #3}
\def\AdM #1 #2 #3{{\em Advances in Math. \/} {\bf #1} (#2) #3}
\def\RMaP #1 #2 #3{{\em Reports on Math. Phys. \/} {\bf #1} (#2) #3}
\def\IJM #1 #2 #3{{\em Ill. J. Math. \/} {\bf #1} (#2) #3}
\def\APP #1 #2 #3{{\em Acta Phys. Polon. \/} {\bf #1} (#2) #3}
\def\TMP #1 #2 #3{{\em Theor. Mat. Phys. \/} {\bf #1} (#2) #3}
\def\JPA #1 #2 #3{{\em J. Physics \/} {\bf A#1} (#2) #3}
\def\JSM #1 #2 #3{{\em J. Soviet Math. \/} {\bf #1} (#2) #3}
\def\MPLA #1 #2 #3{{\em Mod. Phys. Lett. \/} {\bf A #1} (#2) #3}
\def\JETP #1 #2 #3{{\em Sov. Phys. JETP \/} {\bf #1} (#2) #3}
\def\JETPL #1 #2 #3{{\em  Sov. Phys. JETP Lett. \/} {\bf #1} (#2) #3}
\def\PHSA #1 #2 #3{{\em Physica} {\bf A#1} (#2) #3}
\def\CQG #1 #2 #3{{\em Class. Quantum Grav. \/} {\bf #1} (#2) #3}
\def\SJNP #1 #2 #3{{\em Sov. J. Nucl. Phys. (Yadern.Fiz.) \/} {\bf #1} (#2) #3}
\makeatletter
 \@addtoreset{equation}{section}
\def\theequation{\thesection.\arabic{equation}}
 \makeatother

\begin{document}
\thispagestyle{empty}

\title{
\begin{flushright}
{{\small {DFPD 00/TH/31, ESI-912 \vskip-0.5cm
 hep-th/0007048}}}
\end{flushright}
~\\
{\bf Superembeddings, Partial Supersymmetry Breaking and
Superbranes}
 }

\bigskip
\author{
 Paolo Pasti, Dmitri Sorokin\thanks{On leave of absence from
Institute for Theoretical Physics, NSC Kharkov Institute of
Physics and Technology, 61108 Kharkov, Ukraine}~~and
Mario Tonin\\
~\\
{\it Universit\`a Degli Studi Di Padova,
Dipartimento Di Fisica ``Galileo Galilei''}\\
{\it ed INFN, Sezione Di Padova Via F. Marzolo, 8, 35131 Padova,
Italia}\\
e-mail: pasti,sorokin,tonin@pd.infn.it
}
\date{}

\maketitle

\abstract{
It is advocated that the superembedding approach is a generic
covariant method for the description of superbranes as models of
(partial) spontaneous supersymmetry breaking. As an illustration
we construct (in the framework of superembeddings) an
$n=1$, $d=3$ worldvolume superfield action for a supermembrane
propagating in $N=1$, $D=4,5,7$ and 11--dimensional supergravity
backgrounds. We then show how in the case of an
$N=1$, $D=4$ target superspace gauge fixing local worldvolume
superdiffeomorphisms in the covariant supermembrane action
results in an effective $N=2$, $d=3$ supersymmetric field theory
with $N=2$ supersymmetry being spontaneously broken down to
$N=1$. The broken part of $N=2$, $d=3$ supersymmetry is nonlinearly
realized when acting on Goldstone $N=1$, $d=3$ superfields, which
describe physical degrees of freedom of the model. As an
introduction to the formalism, the procedure of getting effective
field theories with partially broken supersymmetry  by gauge
fixing covariant superbrane actions is also demonstrated with a
simpler example of a massive $N=2$, $D=2$ superparticle.}

\newpage
\section{Introduction}

Superbranes are extended relativistic objects which arise as
solitons of supersymmetric field theories. The dynamics of brane
fluctuations can in turn be effectively described by quantum field
theories on the worldvolumes of the superbranes. This is a
manifestation of various kinds of dualities which have been found
and extensively exploited in string/M--theory to gain deeper
insight into its nonperturbative quantum structure. Interest in
effective field theories on branes is also caused by their
relevance to brane--world realizations of the Universe considered
recently. For this it is useful to have an explicit form of brane
effective actions, which generically are supersymmetric.

Single branes usually preserve half of target--space
supersymmetry associated with supertrans\-lations along the
worldvolume. So the fluctuations of such superbranes are
described by supersym\-metric worldvolume sigma--models. In the
standard "Green--Schwarz" formulation of superbrane dynamics
supersymmetry on the brane arises upon gauge fixing worldvolume
diffeomorphisms and a local worldvolume fermionic
$\kappa$--symmetry. Such gauge fixing breaks target superspace
covariance, and supersymmetric transformations associated with
the directions transverse to the brane become nonlinearly
realized on the physical modes of brane fluctuations. These brane
fluctuations can be interpreted as Goldstone modes of the
spontaneously broken (nonlinearly realized) supertranslation
symmetries of the target superspace. Note that from the point of
view of the observer ``living" on the brane it is half of the
worldvolume (space--time) supersymmetry which is spontaneously
broken. Thus, superbranes provide us with a mechanism of partial
spontaneous breaking of space--time supersymmetry
\cite{polch,achu}, the resulting supersymmetric worldvolume
non--linear sigma--models are known to be of the Volkov--Akulov
type \cite{va}.

Superbranes as models of (partial) spontaneous supersymmetry
breaking have been under study for more than a dozen of years.
One of the methods used for their description
\cite{bg2}--\cite{west} has been the group--theoretical (coset
space) method of nonlinear realizations of spontaneously broken
symmetries \cite{ccwz,volkov}. In this formulation the superbrane
dynamics is from the beginning described in a physical (or
``static") gauge where all pure gauge degrees of freedom are
eliminated and only worldvolume fields corresponding to the brane
physical modes remain. The physical modes form a supermultiplet
of unbroken worldvolume supersymmetry, and thus the dynamics of
these modes can be formulated in terms of worldvolume (Goldstone)
superfields at least on the mass shell. In some cases, such as an
$N=2$, $D=4$  Dirichlet 3-brane \cite{bg2} and an N=1, D=4
supermembrane \cite{ik1} \footnote{The number $N$ of the
supersymmetries stands for the number of irreducible spinor
supercharges.}, one can also construct worldvolume superfield
actions describing their off--shell dynamics. We should note that
in the method of nonlinear realizations a systematic way of
constructing gauge fixed superbrane actions with the use of
Goldstone superfields is still lacking, though the actions
written in the components of the Goldstone supermultiplet are
well known. These are the Green--Schwarz--type brane actions in
the physical gauge. To obtain the superfield action for the field
theory with partially broken supersymmetry one passes from the
method of nonlinear realizations to a method which can be
conventionally called the method of ``linear" realization
\cite{ik1} of spontaneously broken supersymmetry. Different, though
related, recipes have been proposed to construct superfield
actions in the framework of the latter approach \cite{bg2,rt}
(see also \cite{kuz} for relevant duality--symmetric
constructions).

To describe models with partial supersymmetry breaking the methods
of nonlinear and linear realizations operate with {\it a priori}
different Goldstone superfields. These superfields are usually
related to each other through complicated expressions (see
\cite{ik1} for the example of the N=1, D=4 supermembrane), so
that even if a superfield action is known in the linear
realization method, in general, it is difficult to rewrite it in
terms of the Goldstone superfields of the nonlinear realization
approach, which upon integration over the Grassmann--odd
coordinates should directly yield the gauge fixed Green--Schwarz
action. As a result a direct relationship of the existing
Goldstone superfield actions with the Green--Schwarz formulation
of superbranes has not been established yet. Such a relationship
has only been checked for the bosonic sectors of the actions,
which were shown to coincide either with the gauge fixed
Nambu--Goto or Dirac--Born--Infeld action depending on the type
of the superbrane considered
\cite{ik1,bg2} \footnote{The $N=1$ supersymmetric Dirac--Born--Infeld
action was first constructed in \cite{fer}.}, while  the fermionic
sectors of different formulations can in general be related by a
highly nontrivial redefinition of the fermionic fields.

A limitation of the methods of partial spontaneous supersymmetry
breaking is that they are suitable for the description of
superbranes propagating in superbackgrounds invariant under global
supersymmetry (or, in other words, in target superspaces with
isometries). The Goldstone superfield actions proposed by now
describe superbranes in flat superbackgrounds with the isometries
generated by super--Poincare algebras.  Goldstone superfield
actions for superbranes in more complicated superbackgrounds with
isometries, such as supersymmetric anti--de--Sitter
configura\-tions of multidimensional supergravities have not been
constructed yet, though component gauge fixed superbrane actions
in $AdS$ superbackgrounds have been intensively studied
\cite{2bs}--\cite{zhou} in connection with the $AdS/CFT$
correspondence conjecture \cite{ads}.

A geometrical approach which describes the dynamics of the
superbranes in arbitrary super\-gravity backgrounds is the method
of superembeddings. This is a generic method for formulating the
theory of superbranes. Other known superbrane formulations
(including the method of nonlinear realizations) follow from the
superembedding approach (see \cite{s} for a recent review).

Superembedding is an elegant and geometrically profound
formulation which is based on a supersymmetric extension of the
classical surface theory applied to the description of superbrane
dynamics by means of embedding worldvolume supersurfaces into
target superspaces \cite{bpstv,hs1,hs2}. Thus, this approach is
manifestly supersymmetric and covariant both on the
superworldvolume and in target superspace. The fermionic
$\kappa$--symmetry of the Green--Schwarz formulation has its
origin in local worldvolume supersymmetry \cite{STV}.

For superembedding to be relevant to the description of
superbranes it should be specified by imposing an appropriate
embedding condition. This condition has a clear geometrical
meaning. Let us consider a supersurface ${\cal M}$ parametrized by
$d=p+1$ bosonic coordinates $\xi^m$ and ${\mathbf n}$ fermionic
coordinates $\eta^\mu$, which we will collectively call
\begin{equation}\label{z}
z^M=(\xi^m,~\eta^\mu), \quad m=0,1,\dots ,p, \quad \mu=1,\dots ,n.
\end{equation}

The geometry of the supersurface is described in a
superdiffeomorphism invariant way by a set of supervielbein
one--forms
\begin{equation}\label{e}
e^A(z)=dz^Me^{~A}_M=\left(e^a(\xi,\eta),~e^\alpha(\xi,\eta)\right),
\end{equation}
which form a local basis in the cotangent space of ${\cal M}$. The
indices $a$ and $\alpha$ are, respectively, the indices of the
vector and a spinor representation of the group $SO(1,p)$ of local
rotations in the cotangent space. The indices $\alpha$ are (in
general) cumulative in the sense that they also include indices of
a group $SO(D-p-1)$ which is the group of internal automorphisms
of the Grassman--odd subspace of ${\cal M}$ possessing $n=D-p-1$
extended supersymmetry.

 Let us now embed this supersurface into a curved target
superspace
$\underline M$ parametrized by $D$ bosonic coordinates
$X^{\underline m}$ and {\bf 2n} fermionic coordinates
$\Theta^{\underline\mu}$, which we will collectively call
\begin{equation}\label{Z}
Z^{\underline M}=(X^{\underline m},~\Theta^{\underline\mu}), \quad
\underline m=0,1,\dots,D-1, \quad \underline\mu=1,\dots,2n.
\end{equation}
Note that for embedding we have chosen a supersurface with the
number of Grassmann--odd directions being half the number of
target--superspace Grassmann--odd directions. This is for being
able to identify $n$ local worldvolume supersymmetries with $n$
independent fermionic $\kappa$--symmetries of the standard
(Green--Schwarz) formulation of superbrane dynamics. In this
paper we shall also deal with supersurfaces with a less number of
fermionic coordinates.

The geometry of the target superspace is described in a
superdiffeomorphism invariant way by a set of supervielbein
one--forms
\begin{equation}\label{E}
E^{\underline A}(Z)=dZ^{\underline M}E^{~\underline A}_{\underline
M}=\left( E^{\underline a}(X,\Theta), E^{\underline
\alpha}(X,\Theta)\right),
\end{equation}
which form a local frame in the cotangent space of the target
superspace. The indices $\underline a$ and $\underline\alpha$ are,
respectively, the indices of the vector and a spinor
representation of the group $SO(1,D-1)$ of local rotations in the
$\underline M$ cotangent space.

Superembedding is a map of ${\cal M}$ into $\underline M$ which is
locally described by $X^{\underline m}$ and
$\Theta^{\underline\mu}$ as functions of the supersurface
coordinates
\begin{equation}\label{map}
z^M\quad \rightarrow \quad Z^{\underline M}(z)=\left(X^{\underline
m}(\xi, \eta),~\Theta^{\underline\mu}(\xi,\eta)\right).
\end{equation}
 The map induces the pullback onto the supersurface of the
target superspace one--form \p{E}. In particular, the vector
supervielbein $E^{\underline a}$ pullback is a one--superform on
the supersurface. It has the following decomposition in the local
basis \p{e} on $\cal M$
\begin{equation}\label{pull}
E^{\underline a}(z)=e^a(z)E^{~\underline
a}_a(Z(z))+e^\alpha(z)E^{~\underline a}_\alpha(Z(z)).
\end{equation}
The superembedding condition we are interested in is the
vanishing of the worldvolume spinor components of $E^{\underline
a}(z)$
\begin{equation}\label{emb}
E^{~\underline a}_\alpha(Z(z))=0.
\end{equation}
In other words eq. \p{emb} is a superfield constraint on \p{map}
which singles out the superembeddings such that the pullback of
the supervielbein $E^{\underline a}$ has non--zero components
only along vector directions of the supersurface. It can be shown
that \p{emb} sets an induced supergeometry on the embedded
supersurface \cite{bpstv}, i.e. that the worldvolume supervielbein
\p{e} is completely determined in terms of the components of the
target space supervielbein pullback \p{E}. This is in accordance
with a well known fact that no supergravity propagates on the
superbrane.

Thus, in the superembedding approach superbrane dynamics is
described in the framework of a worldvolume superfield formalism.

Eq. \p{emb} is the basic superembedding condition for the
description of all superbranes. In some cases the superembedding
condition produces only ``kinematic" constraints (such as, for
instance, Virasoro conditions) and does not put superbrane
dynamics on the mass shell. (Examples are $N=1$, $D=2,3,4,6$,10
superparticles \cite{STV}--\cite{ps92} and heterotic superstrings
\cite{hsstr}--\cite{dghs92}). In these cases several methods have
been developed \cite{STV,to,ds,gs92,dghs92} for constructing
worldvolume superfield actions which produce dynamical equations
of motion of the superbranes. Alternatively, the dynamical
equations of motion can be obtained from a supersymmetric
generalization of the condition of minimal area embedding imposed
on the second fundamental form of the supersurface \cite{bpstv}.

In other cases, such as the M--theory branes (a $D=11$
supermembrane \cite{gsn2,bpstv} and a super-5--brane \cite{hs2}),
the superembedding condition contains all information about the
classical dynamics of the superbranes (i.e. the constraints and
the equations of motion). In these cases the worldvolume
superfield actions have not been found, and one should instead
deal with generalized action functionals \cite{bsv,hrs}, or
conventional Green--Schwarz--like actions.

Thus, the superembedding approach provides systematic geometrical
methods for getting worldvolume superfield equations of motion,
and for constructing worldvolume superfield brane actions when the
superembedding condition is off the mass shell.

The knowledge of worldvolume superfield actions for superbranes
in the covariant superembedding approach can be used to derive
corresponding gauge fixed superbrane actions in terms of
Goldstone superfields, which describe field theories with partial
spontaneous supersymmetry breaking in the method of nonlinear
realizations. The general procedure is as follows. One chooses the
superbackground to be a superspace with isometries and studies
spontaneous breaking of the isometries when a superbrane
propagates in this superbackground.

The simplest case is when the target superspace \p{Z}, \p{E} is
flat. Then one deals with a global $N=2n$ supersymmetry in the
target superspace broken down to its $n$-extended subsupergroup.
This subsupergroup is associated with $n$ worldvolume
superdiffeomorphisms which reduce, upon imposing a physical gauge
condition, to an $n$-extended (unbroken) global supersymmetry on
the superworldvolume. The physical gauge condition identifies the
supercoordinates \p{z} of the superworldvolume with a part of the
target superspace coordinates \p{Z}, \p{map}
\begin{equation}\label{static}
X^m(\xi,\eta)=\xi^m, \quad \Theta^{\mu}(\xi,\eta)=\eta^\mu, \quad
m=0,1,\dots,p, \quad \mu=1,\dots,n.
\end{equation}
Using the worldvolume superdiffeomorphisms $z^M~\rightarrow~\hat
z^M(z)$ it is always possible, at least locally, to make such a
choice of the target superspace coordinates on the superbrane.

The remaining worldvolume superfields \p{map}
\begin{equation}\label{gold}
X^{i'}(\xi,\eta),  \quad \Theta^{\mu'}(\xi,\eta), \quad
i'=p+1,\dots,D-1, \quad \mu'=n+1,\dots,2n
\end{equation}
are the Goldstone superfields associated with spontaneously broken
supertranslations of the superbackground along the bosonic and
fermionic directions normal to the brane superworldvolume. They
describe the transverse fluctuations of the superbrane and
transform nonlinearly under broken supersymmetry.

The superfields \p{gold} are not independent. They are related to
each other by the superembed\-ding condition \p{emb} which now
plays the role of a condition ensuring a so called ``inverse
Higgs" effect \cite{ivanov}, i.e. when the number of Goldstone
fields gets reduced by making some of them dependent on the other
ones. The similarity of the superembedding condition and the
inverse Higgs constraint has been known for a long time
\cite{vz,stvz} \footnote{This similarity was pointed out to authors of \cite{vz,stvz}
by I. Bandos.}. Though, as we have discussed above, the former has
a much more general and profound geometrical meaning.

In this paper we illustrate the general procedure of passing from
the superembedding approach to the method of nonlinear
realizations with instructive examples of a massive superparticle
in $N=2$, $D=2$ superspace and of a supermembrane in $N=1$, $D=4$
superspace. The paper may be regarded as an up--to--date revision
and generalization of the results of the study of the $D=2$
superparticle and the $D=4$ supermembrane considered in
references \cite{achu,ik,gaunt,hrs,ik1}.

In Section 3 we present a new simple, worldvolume and target
space supersymmetric, form of the action which describes the
dynamics of a supermembrane in a superbackground of any dimension
where the supermembrane is allowed to propagate by the brane scan
\cite{scan}, for instance, in backgrounds of four-- and
eleven--dimensional supergravity. This action can be regarded as
a dynamical realization of the superembedding approach. It is
constructed with the use of the worldvolume superfields \p{e} and
\p{map}, and pullbacks of differential forms describing
corresponding supergravity backgrounds. The action possesses
interesting features. For instance, its main term is a
superworldvolume integral of the co--dimension two component of a
Wess--Zumino three--form, and it is invariant under super Weyl
transformations of the worldvolume supervielbein
\p{e}. Remember that, in contrast to strings, the
Howe--Tucker--Polyakov formulation of membrane dynamics is not
invariant under Weyl rescaling of the intrinsic worldvolume
metric. In our case the super Weyl symmetry is required for the
superembedding condition to identify intrinsic worldvolume
supergeometry with supergeometry induced by embed\-ding.

We shall demonstrate how the superembedding action is related to
the Green--Schwarz--type formulation of \cite{bst1} and
\cite{hrs}, and how in the case of an $N=1$, $D=4$ flat target
superspace it reduces, in the physical gauge \p{static}, to a
superfield generalization of the component action of \cite{achu}.
The Goldstone superfield action thus obtained describes an $N=2$,
$d=3$ dimensional supersymmetric theory of a self--interacting
scalar supermultiplet with one linearly realized supersymmetry and
another one being spontaneously broken. The latter is realized as
a nonlinear transformation of a single
$N=1$, $d=3$ Goldstone scalar superfield. In this way we get the
action for the Goldstone superfield in the method of nonlinear
realizations, which is thus directly related to the
superembedding approach and to the Green--Schwarz formulation. We
also demonstrate the relationship of this supermembrane action to
the action constructed within the framework of ``linear"
realizations \cite{ik1}.

Section 2 of the paper is devoted to a detailed consideration of
a simpler example of a one--dimensional sigma--model  with
partially broken
$N=2$ supersymmetry which is obtained from the dynamics of a
massive superparticle in an $N=2$, $D=2$ target superspace. This
section may be regarded as an introduction into the formalism and
as an illustration of the links between the superembedding
approach and the methods of spontaneous supersymmetry breaking.
This should simplify understanding the example of the
supermembrane considered in Section 3. In Conclusion we discuss
open problems and outlook.

\section{The massive $N=2$, $D=2$ superparticle}
In the framework of the superembedding approach the massive
superparticle in an $N=2$, $D=2$ target superspace has been
studied in \cite{ik,gaunt,ps92}. ($N=2$ here stands for the number
of one--component $D=2$ Majorana--Weyl spinors which form a
two--component Majorana spinor.) We start the consideration from
an action considered in
\cite{ik,gaunt}, and then generalize it in an appropriate way for
being able to impose the physical gauge
\p{static} discussed in the Introduction.

In the case of a flat target superspace the action has the
following form\footnote{One can compare this action with the
standard massive superparticle action presented in eq. \p{com1}
of the Appendix. The $\eta=0$ components of the superfields
$P_{\underline a}$, $X^{\underline a}$ and
$\Theta^{\underline\alpha}$ correspond to the variables
$p_{\underline a}$, $x^{\underline a}$ and $\theta^{\underline
\alpha}$ of \p{com1}.}
\begin{equation}\label{2n2d}
S=\int d\tau d\eta \left [iP_{\underline a}(DX^{\underline
a}-iD\bar\Theta\Gamma^{\underline a}\Theta)-m
D\bar\Theta\Gamma^{2}\Theta \right],
\end{equation}
where the superworldline ${\cal M}$ of a particle of mass $m$ is
parametrized by the bosonic time variable $\tau$ and the real
fermionic variable $\eta$, and
\begin{equation}\label{D}
D={\partial\over{\partial\eta}}+i\eta\partial_\tau \quad \
D^2={1\over 2}\{D,D\}=i\partial_\tau
\end{equation}
 is a `flat' Grassmann
covariant derivative. The image of ${\cal M}$ in the target
superspace is described by the scalar worldvolume superfields
$X^{\underline a}(\tau,\eta)$, ($a=0,1$) and
$\Theta^{\underline\alpha}(\tau,\eta)$, ($\alpha=1,2$), which
transform as a vector and a Majorana spinor
($\bar\Theta=\Theta^TC$) under the action of the $D=2$ Lorentz
group $SO(1,1)$. The $D=2$ Dirac matrices $\Gamma^{\underline a}$,
$\Gamma^{2}$ and the charge conjugation matrix
$C_{\underline\alpha\underline\beta}$ are chosen to be in a
Majorana representation
\begin{equation}\label{Gamma}
(\Gamma^{0})^{\underline\alpha}_{~~\underline\beta}= \left(
\begin{array}{cc}
0  &  -1\\
1 &  0
\end{array}
\right)\,, \quad
(\Gamma^{1})^{\underline\alpha}_{~~\underline\beta}= \left(
\begin{array}{cc}
-1  &  0\\
0 &  1
\end{array}
\right)\, , \quad \Gamma^2=\Gamma^0\Gamma^1=\left(
\begin{array}{cc}
0  &  -1\\
-1  &   0
\end{array}
\right) \,.
\end{equation}
\begin{equation}\label{C}
 C_{\underline\alpha\underline\beta}=C^{\underline\alpha\underline\beta}=\left(
\begin{array}{cc}
0  &  1\\
-1 &  0
\end{array}
\right)\,.
\end{equation}
\begin{equation}\label{Gamma1}
\Gamma^{0}_{\underline\alpha\underline\beta}= \left(
\begin{array}{cc}
1  &  0\\
0 &  1
\end{array}
\right)\,, \quad \Gamma^{1}_{\underline\alpha\underline\beta}=
\left(
\begin{array}{cc}
0  &  1\\
1 &  0
\end{array}
\right)\, , \quad \Gamma^2_{\underline\alpha\underline\beta}=\left(
\begin{array}{cc}
-1   &  0\\
0   &   1
\end{array}
\right) \,.
\end{equation}
The rules of raising and lowering the spinor indices are
$\Theta_{\underline\alpha}=C_{\underline\alpha\underline\beta}\Theta^{\underline\beta}$\,,
$\Theta^{\underline\alpha}=\Theta_{\underline\beta}C^{\underline\beta\underline\alpha}$.

The action \p{2n2d} is invariant (up to a total derivative) under
the global supersymmetry transformations in the target superspace
(note that the second term in \p{2n2d} is of a Wess--Zumino type)
\begin{equation}\label{tsusy}
\delta\Theta^{\underline{\alpha}}=\epsilon^{\underline{\alpha}},
\qquad \delta X^{\underline{a}}=
i\bar\Theta\Gamma^{\underline{a}}\delta\Theta,
\end{equation}
and under local transformations of the worldvolume coordinates of
the following form
\begin{equation}\label{wsusy}
\begin{array}{rl}
\tau'-\tau&=\delta\tau=2\Lambda(\tau,\eta)-\eta D\Lambda,
\\
\eta'-\eta&=\delta \eta = -iD\Lambda,\\
D'-D&=\delta D=-\partial_\tau\Lambda D,
\end{array}
\end{equation}
where $\Lambda(\tau,\eta)=a(\tau)+i\eta\alpha(\tau)$ is the
superreparametrization parameter which contains the worldline
bosonic reparametrization parameter $a(\tau)$ and the local
supersymmetry parameter $\alpha(\tau)$.

Under \p{wsusy} the superfields $X^{\underline a}(\tau,\eta)$ and
$\Theta^{\underline\alpha}(\tau,\eta)$ transform as scalars, and
the superfield $P_{\underline a}(\tau,\eta)$ in the first term of
\p{2n2d} transforms in an appropriate way to ensure the invariance
of the action. Note that to ensure the invariance of the action
\p{2n2d} under \p{wsusy} it was not necessary to introduce the
worldvolume supervielbein \p{e}. However, we shall need it later
for promoting local worldvolume supersymmetry \p{wsusy} to general
superdiffeomorphisms, which is required for imposing the physical
gauge condition \p{static}.

$P_{\underline a}$ is the Lagrange multiplier\footnote{The
leading component $P_{\underline a}|_{\eta=0}$ of $P_{\underline
a}(\tau,\eta)$ is the particle canonical momentum.} whose
variation in \p{2n2d} produces the superembedding condition
\begin{equation}\label{emb1}
DX^{\underline a}-iD\bar\Theta\Gamma^{\underline a}\Theta=0.
\end{equation}
Eq. \p{emb1} is a flat target space counterpart of the condition
\p{emb} where now $E^{\underline a}=dX^{\underline
a}-id\bar\Theta\Gamma^{\underline a}\Theta$ and $e^a$ and
$e^\alpha$ are, respectively, $d\tau-id\eta \eta$  and $d\eta$.

In the case under consideration eq. \p{emb1} is a constraint
which relates the superfields $X^{\underline a}(\tau,\eta)$ and
$\Theta^{\underline\alpha}(\tau,\eta)$. It produces the
relativistic energy--momentum condition $P_{\underline
a}P^{\underline a}|_{\eta=0}=m^2$, but does not contain dynamical
equations of motion \cite{STV,ik,gaunt}. The latter are derived
by varying the action \p{2n2d} with respect to $X^{\underline a}$
and $\Theta^{\underline\alpha}$.

Our goal is to gauge fix the local superreparametrizations of the
superworldline, to solve eq. \p{emb1} explicitly in terms of an
independent superfield and to substitute this solution into the
second (Wess--Zumino) term of the action \p{2n2d}. The resulting
action will describe a one--dimensional supersymmetric nonlinear
sigma--model with partially broken $N=2$ supersymmetry.

We would like to relate the superembedding formulation of the
superparticle to the Goldstone superfield action of \cite{achu},
which up to a normalization is
\begin{equation}\label{achu}
S=im\int d\tau d\eta{{\partial_\tau\Phi D\Phi}
\over{1+\sqrt{1-(\partial_\tau\Phi)^2}}}.
\end{equation}
$\Phi(\tau,\eta)$ is a scalar superfield which describes the
superparticle physical degrees of freedom in the physical gauge.
Hence, we should impose the condition \p{static}. In the case
under consideration it takes the form
\begin{equation}\label{static1}
X^0(\tau,\eta)=\tau, \quad \Theta^1(\tau,\eta)=\eta.
\end{equation}
To be able to impose this condition which gauge fixes {\it two}
superfield variables $X^0$ and $\Theta^1$ one must have
worldvolume superreparametrizations with {\it two} independent
superfield parameters. However, in the form \p{2n2d} the
superparticle action is invariant under the one--parameter
transformations \p{wsusy}. As we show in the Appendix, the
one--parameter superreparametrizations can be used to impose a
light--cone gauge condition $X^-(\tau,\eta)\equiv
(X^0-X^1)=\tau$, but not the conditions \p{static1}. Therefore,
we should modify the action \p{2n2d} in such away that it becomes
invariant under general superdiffeomorphisms of the worldline
supersurface
\begin{equation}\label{dif}
\tau'=\tau'(\tau,\eta),\quad \eta'=\eta'(\tau,\eta)
\end{equation}
 characterized
by two independent superfunctions. For this we should
``covariantize" the action \p{2n2d}, i.e. couple it to worldline
supergravity by introuducing the worldline supervielbein \p{e}
\begin{equation}\label{e1}
e^A=dz^Me^{~A}_M, \quad z^M=(\tau,\eta).
\end{equation}
It has been proved convenient to choose the matrix $e^{~A}_M$ in
the form
\begin{equation}\label{e11}
e^{~A}_{M}= \left(
\begin{array}{cc}
e &  -ife\\
i\tilde fe   &  \tilde e
\end{array}
\right)\,,
\end{equation}
then its inverse is
\begin{equation}\label{e-1}
\quad e^{~M}_{A}=\left(
\begin{array}{cc}
e^{-1}(1+{{ef\tilde f}\over{\tilde  e}}) &  {{if}\over {\tilde e}}\\
{{-i\tilde f}\over{\tilde e}}   &  \tilde e^{-1}(1+{{e\tilde f
f}\over{\tilde  e}})
\end{array}
\right)\, ,
\end{equation}
where $e(\tau,\eta)$ and $\tilde e(\tau,\eta)$ are bosonic
 and $f(\tau,\eta)$ and $\tilde
f(\tau,\eta)$ are fermionic worldvolume superfields.

It is straightforward to ``covariantize" the first term of the
action \p{2n2d} by replacing the flat covariant derivative $D$
with its curved counterpart
\begin{equation}\label{calD}
{\cal D}=e^{~M}_\eta\partial_M={1\over \tilde e}\left(
(1+{{e\tilde ff}\over{\tilde  e}})
\partial_\eta+{if}\partial_\tau\right),
\end{equation}
where $e^{~M}_\eta$ is the second column of the supervielbein
matrix  \p{e-1}.

 However, as far as the second term in \p{2n2d} is concerned,
its generalization is more subtle. It must not spoil the property
of this term to be of the Wess--Zumino type, i.e to be invariant
under target--space supersymmetry \p{tsusy} up to a total
derivative.

To find the appropriate generalization of the Wess--Zumino term
we first consider the super\-em\-bed\-ding action for a massless
superparticle in an $N=1$, $D=3$ superspace \cite{STV} in the form
invariant under \p{dif} and then perform its dimensional
reduction to the massive $N=2$, $D=2$ superparticle action (as in
\cite{ps92}).

The $N=1$, $D=3$ massless superparticle action in question has the
following form
 $$ S=i\int d\tau d\eta P_{\underline {\hat
a}}({\cal D}X^{\underline {\hat a}}-i{\cal
D}\bar\Theta\Gamma^{\underline {\hat a}}\Theta)
 $$
 \begin{equation}\label{d3}
=i\int d\tau d\eta P_{\underline a}({\cal D}X^{\underline
a}-i{\cal D}\bar\Theta\Gamma^{\underline a}\Theta)+i\int d\tau
d\eta P_2({\cal D}X^2-i{\cal D}\bar\Theta\Gamma^2\Theta),
\end{equation}
where $X^{\underline {\hat a}}$ $(\hat a=0,1,2)$ are bosonic
coordinates of the $D=3$ superspace, and the covariant fermionic
derivative $\cal D$ has been introduced in \p{calD}.

By construction the action \p{d3} is invariant under the
target--space supersymmetry trans\-formations
\begin{equation}\label{SUSY}
\delta\Theta^{\underline{\alpha}}=\epsilon^{\underline{\alpha}},
\qquad \delta X^{\underline{\hat a}}=
i\bar\Theta\Gamma^{\underline{\hat a}}\delta\Theta,
\end{equation}
and under the worldvolume superdiffeomorphisms \p{dif} provided
the Lagrange multiplier super\-field $P_{\underline{\hat
a}}(\tau,\eta)$ (whose leading component is associated with the
superparticle canonical momentum \cite{STV}) transforms in an
appropriate way.

In addition, since only half of the components of the inverse
supervielbein \p{e-1} enter the eq. \p{d3} through the covariant
derivative ${\cal D}$, the action is invariant under super--Weyl
transformations
\begin{eqnarray}\label{sw}
&&\delta_fe^{~M}_\tau=\phi_f(\tau,\eta)e^{~M}_\eta, \qquad
\delta_f e^{~M}_\eta=0, \qquad \delta_fP_{\underline{\hat
a}}=0;\nonumber\\
 &&\delta_b
e^{~M}_\eta=\phi_b(\eta,\tau)e^{~M}_\eta,\qquad
 \delta_be^{~M}_\tau=0,
\qquad \delta_bP_{\underline{\hat
a}}=-\phi_b(\eta,\tau)P_{\underline{\hat a}},
\end{eqnarray}
where $e^{~M}_\tau$ and $e^{~M}_\eta$ are, respectively, the first
and the second column of \p{e-1}, and $\phi_f(\tau,\eta)$ and
$\phi_b(\tau,\eta)$ are a bosonic and fermionic parameter of the
super--Weyl transformations.

The transformations \p{sw} can be used to put $\tilde f=0$ and
$\tilde e=1$ in \p{e11} and \p{e-1}, which then reduce to
\begin{equation}\label{e11r}
e^{~A}_{M}= \left(
\begin{array}{cc}
e &  -ife\\
0  &  1
\end{array}
\right)\,
\end{equation}
\begin{equation}\label{e-1r}
\quad e^{~M}_{A}=\left(
\begin{array}{cc}
e^{-1} &  {if}\\
0 &  1
\end{array}
\right)\, ,
\end{equation}
and the covariant derivative in \p{d3} takes the form
\begin{equation}\label{calDr}
{\cal D}=\partial_\eta+if(\tau,\eta)\partial_\tau, \quad {\cal
D}^2=i({\cal D}f)\partial_\tau.
\end{equation}
The choice of the worldvolume supervielbein in the form
\p{e11r}--\p{calDr} fixes the super--Weyl invariance of the action
\p{d3}, and in what follows we shall work in this gauge.

The dimensional reduction of the action \p{d3} down to the $N=2$,
$D=2$ superparticle action is carried out in the following way.
The space dimension associated with the coordinate $X^2$ is
assumed to be compactified on a circle, and the superparticle is
restricted to move along the circle with a constant momentum
whose value determines the particle mass in an effective
(uncompactified) two--dimensional space--time. Technically this is
done by solving for the equation of motion of $X^2(\tau,\eta)$,
which is
\begin{equation}\label{x2}
{\cal D}P_2+iP_2\partial_\tau f=0,
\end{equation}
and substituting the solution back into the action \p{d3}.

The general solution of \p{x2} is
\begin{equation}\label{x2s}
P_2=-{m\over{{\cal D}f}},
\end{equation}
which can be checked using the properties of the covariant
derivative \p{calDr}. In \p{x2s} $m$ is a constant mass parameter.

Substituting \p{x2s} into eq. \p{d3} and noticing that the term
with ${\cal D}X^2$ becomes a total derivative, we arrive at the
desired form of the $N=2$, $D=2$ superparticle action
\begin{equation}\label{d2}
S=i\int d\tau d\eta P_{\underline a}({\cal D}X^{\underline
a}-i{\cal D}\bar\Theta\Gamma^{\underline a}\Theta)-\int d\tau
d\eta {m\over{{\cal D}f}}{\cal D}\bar\Theta\Gamma^2\Theta.
\end{equation}

We are now in a position to use the worldvolume
superdiffeomorphisms \p{dif} for imposing the physical gauge
\p{static1}. Then the time and space component of the
superembedding condition \p{emb1} reduce, respectively, to the
following equations (which are obtained using the explicit form
of the Dirac matrices \p{Gamma})
\begin{equation}\label{0}
if(\tau,\eta)=i\eta+i\Psi {\cal D} \Psi,
\end{equation}
\begin{equation}\label{1}
{\cal D}\Phi=2i\Psi,
\end{equation}
where
 \begin{equation}\label{P} \Psi(\tau,\eta)\equiv\Theta^2,
\qquad \Phi(\tau,\eta)\equiv X^1+i\eta\Psi.
\end{equation}

From the equation \p{0} we find the expression for the worldvolume
supervielbein component $f(\tau,\eta)$ in terms of the ``matter"
superfield $\Psi(\tau,\eta)$
\begin{equation}\label{f}
f={{\eta+\Psi\partial_\eta\Psi}\over{1+i\Psi\partial_\tau\Psi}}=\eta+\Psi
D\Psi, \quad D=\partial_\eta+i\eta\partial_\tau.
\end{equation}
Then we can substitute \p{f} into \p{1} and solve this equation
using a nice trick of Bagger and Galperin \cite{bg2} (which we
describe in the Appendix). We thus find the expression for
$\Psi(\tau,\eta)$ in terms of the unconstrained superfield
$\Phi(\tau,\eta)=\varphi(\tau)+i\eta\psi(\tau)$, which describes
the physical bosonic and fermionic degrees of freedom of the
superparticle
\begin{equation}\label{psi}
\Psi=-{{iD\Phi}\over{1+\sqrt{1-(\partial_\tau \Phi)^2}}}.
\end{equation}
From the form of eqs. \p{1} and \p{f} one can see why the physical
gauge \p{static} was inadmissible in the case of the
superparticle action \p{2n2d}, where $f=\eta$. Putting $f=\eta$
already fixes the Grassmann--odd part of the diffeomorphisms
\p{dif}. So if in addition we put also an extra condition
$\eta=\Theta^1$, from \p{f} it would follow that $\Psi D\Psi=0$
and $\Psi=D\Phi$. Then, as can be easily checked, the equation of
motion of $\Psi$ derived from \p{2n2d} would reduce to $D\Psi=0$,
and thus result in $\partial_\tau\Phi=0$, which is too
restrictive, since it describes a ``static" particle (recall that
particle motion in space is governed by second order differential
equations).

Since we have explicitly solved the superembedding condition
\p{emb1} in the physical gauge \p{static1}, the action \p{d2},
which now contains only the second (Wess--Zumino) term, reduces
to the following form
\begin{equation}\label{d2gauge}
S=\int d\tau d\eta {m\over{{\cal D}f}}(\eta-\Psi{D}\Psi).
\end{equation}

Upon some algebraic manipulations one can show that
\begin{equation}\label{Df}
{1\over{{\cal D}f}}=1-D{{\Psi D\Psi}\over{1+(D\Psi)^2}}=1+{i\over
2}D{{D\Phi\partial_\tau\Phi}\over{1+\sqrt{1-(\partial_\tau
\Phi)^2}}}
\end{equation}
and
\begin{equation}\label{psidpsi}
\Psi
D\Psi=-i{{D\Phi\partial_\tau\Phi}\over{(1+\sqrt{1-(\partial_\tau
\Phi)^2})^2}}.
\end{equation}
Substituting \p{Df} and \p{psidpsi} into \p{d2gauge} and
integrating by parts we finally arrive at the action which
coincides with eq. \p{achu} up to the ``cosmological'' term $\int
d\tau~ m$, which was skipped in \cite{achu} in order to normalize
to zero the energy of the ``ground" state $D\Phi(\tau,\eta)=0$.

Let us now analyze the symmetries of the action \p{achu}. The
gauge conditions \p{static1} remain invariant under the
combination of the $N=2$, $D=2$ target space supersymmetry
\p{tsusy} and a global relic of the worldline
superdiffeomorphisms \p{dif} which must be related to \p{tsusy}
as follows
\begin{equation}\label{wsusyg}
\delta\eta=\epsilon^1, \quad
\delta\tau=i\eta\epsilon^1+i\Psi(\eta,\tau)\epsilon^2.
\end{equation}
Under the $\epsilon^1$--transformations the superfield
$\Phi(\tau,\eta)$ entering the action \p{achu} varies as the
scalar superfield
\begin{equation}\label{dphi}
\delta\Phi=-\delta_{_{(\epsilon^1)}} z^M\partial_M\Phi.
\end{equation}
Hence, the action \p{achu} is manifestly invariant under $N=1$
global supersymmetry transformations in the worldline superspace
$(\tau,\eta$) associated with the parameter $\epsilon^1$. This is
the supersymmetry which remains unbroken.

The action is also invariant under the second, nonlinearly
realized (and hence spontaneously broken) supersymmetry associated
with the $\epsilon^2$--shifts \p{tsusy} of the superfield
$\Psi(\tau,\eta)=\Theta^2$. Under the
$\epsilon^2$--transformations the Goldstone fermion
 $\Psi$ and its bosonic Goldstone partner $\Phi$ (which is associated with
spontaneously broken translations along the space direction
$X^1$) vary in a nonlinear way
\begin{equation}\label{e2}
\delta\Psi=\epsilon^2(1+i\Psi\partial_\tau\Psi), \quad
\delta\Phi=-i\epsilon^2(2\eta-\Psi\partial_\tau\Phi)=-i\epsilon^2(2\eta-L),
\quad L=\Psi\partial_\tau\Phi={{\partial_\tau\Phi D\Phi}
\over{1+\sqrt{1-(\partial_\tau\Phi)^2}}},
\end{equation}
where $L$ is the Lagrangian density of the action \p{achu}.

The transformations \p{e2} can be easily derived from the
definition \p{P} and \p{psi} of $\Phi$ and $\Psi$, and using their
variation properties with respect to the combination of target
space \p{tsusy} and worldline \p{wsusyg} supersymmetry
transformations with the parameter $\epsilon^2$.

The superfield transformations \p{e2} have been obtained in
\cite{ik2} using somewhat different reasoning course.

We have thus demonstrated how the $N=2$, $D=2$ massive
superparticle action \p{d2} in the doubly supersymmetric
superembedding approach reduces (upon an appropriate gauge fixing
of the local worldvolume superdiffeomorphisms \p{dif}) to the
one--dimensional nonlinear sigma--model \p{achu} exhibiting
partial breaking of $N=2$ global supersymmetry.

In the next section we proceed to the consideration of a more
complicated and interesting example of a three--dimensional field
theory with partially broken supersymmetry describing
supermembrane fluctuations in target superspace.

\section{The supermembrane}
In the framework of the superembedding approach the supermembrane
has been studied in \cite{pt,bpstv,bsv,hrs}. In \cite{gsn2,bpstv}
it has been shown that in a $D=11$ supergravity background the
superembedding condition puts the dynamics of the supermembrane
on the mass shell (i.e. contains the supermembrane equations of
motion) if the worldvolume supersurface \p{z} with $p=2$ has
$n=16$ Grassmann--odd directions, i.e. $N=8$ supersymmetry in
$d=3$ \footnote{The number $N$ of the supersymmetries stands for
the number of Majorana spinor supercharges.}. In this case the
superfield action of the type \p{2n2d} cannot be constructed,
since the Lagrange multipliers $P_{\underline a}$ propagate
redundant degrees of freedom. Instead one can deal with a
generalized action functional \cite{bsv} which, though being not
a fully fledged superworldvolume action, allows one to derive the
superembedding condition and, as a consequence, the full set of
superfield equations of motion of the $D=11$ supermembrane.

The superembedding condition can be relaxed if the worldvolume
supersurface associated with the supermembrane has a less number
of Grassmann--odd directions, for instance, $n=2$. Such an $N=1$,
$d=3$ supersurface is then parametrized by the supercoordinates
\begin{equation}\label{z1}
z^M=(\xi^m,\eta^\mu), \qquad m=0,1,2, \qquad \mu=1,2\,.
\end{equation}
If we embed the supersurface \p{z1} into a $D=4,5,7,11$ target
superspace (with 4,8,16 and 32 Grassmann--odd directions,
respectively) it can be shown, making the analysis described in
\cite{bpstv,hrs}, that the superembedding condition \p{emb} does not
contain the supermembrane equations of motion. Hence, in this case
an $N=1$, $d=3$ superworldvolume action can be constructed for a
supermembrane propagating in a $D=4,5,7$ and 11 supergravity
background (remember that these backgrounds fit into the brane
scan \cite{scan}). Below we give the form of this action.

We should note that the embedding of the supersurface \p{z1} with
only two Grassmann--odd directions into a $D=5,7$ or $D=11$
superbackground does not allow to trade all $\kappa$--symmetries
of the standard formulation \cite{bst1} (e.g. 16 in $D=11$) for
only two supersymmetries of the superworldvolume \p{z1}. In such
a formulation a part of the $\kappa$--transformations remains as a
hidden symmetry. The match of the number of the supersymmetries of
$\cal M$ \p{z1} and the number of $\kappa$--symmetries takes
place when $\cal M$ is embedded into an $N=1$, $D=4$ superspace
with four real Grassmann--odd directions. This last case will be
of our main interest in view of the relationship of the
superembedding approach and the methods of spontaneous
supersymmetry breaking. However, until a certain point we shall
not specify the dimension of the target superspace \p{Z}, \p{E}.

\subsection{The conventional form of the supermembrane action}
The Green--Schwarz--type action for a supermembrane propagating in
an $N=1$, $D=4,5,7$ or $11$ supergravity background has the
following form \cite{bst1}
\begin{equation}\label{2}
S_{M2}=-\int_{{\cal M}_3}d^3\xi\sqrt{-\det{g_{mn}}}-{1\over 2}
\int_{{\cal M}_3}d^3\xi
\varepsilon^{mnp}\partial_mZ^{\underline L}\partial_nZ^{\underline M}
\partial_pZ^{\underline N}A_{\underline{NML}}(Z),
\end{equation}
where
$g_{mn}(\xi)=\partial_mZ^{\underline M}\partial_nZ^{\underline N}(\xi)
E_{\underline M}^{~ \underline a}(Z)
E_{{\underline a}\underline N}(Z)$ is a worldvolume metric
induced by embedding the {\em bosonic} surface
${\cal M}_3$
(parametrized by $\xi^m$) into a curved target superspace \p{Z},
\p{E}.

The supermembrane also minimally couples to a background
three-form superfield $A_{\underline{NML}}(Z)$. In $D=11$ its
leading component $A_{\underline  n\underline m
\underline l}(X)=A_{\underline  n\underline m \underline
l}(Z)|_{\Theta=0}$ is the gauge field of $D=11$ supergravity, and
in $D=5$ it is Hodge dual to a scalar component of the $N=1$,
$D=5$ supergravity multiplet.

In $D=4$ $A_{\underline n\underline m \underline l}(X)$ does not
have any local dynamical degrees of freedom, since its field
strength
$F_{\underline p\underline n\underline m\underline l}=4!\partial_{[\underline p}
A_{\underline n\underline m \underline l]}(X)$ is constant on the
mass shell
\begin{equation}\label{F}
{\cal D}_{\underline p}F^{\underline p\underline n\underline m\underline
l}=0 \quad \rightarrow \quad F^{\underline p\underline
n\underline m\underline l}=c~
\varepsilon^{\underline p\underline n\underline m\underline l}
\end{equation}
Though being locally non--dynamical $F_{\underline p\underline
n\underline m\underline l}$ has a positive energy density and,
hence, contributes to the cosmological constant value. This
mechanism of the dynamical generation of the cosmological
constant has been studied \cite{4form,deal,bousso} as a possible
way of solving the zero cosmological constant problem. We see
that in
$D=4$ the supermembrane naturally couples to such a
``cosmological" field.

The action \p{2} is invariant under target--space
superdiffeomorphisms
\begin{equation}\label{3}
Z'^{\underline M}=Z'^{\underline M}(Z),
\end{equation}
local worldvolume diffeomorphisms
\begin{equation}\label{4}
\xi'^m=\xi'^m(\xi)
\end{equation}
and local fermionic $\kappa$--symmetry transformations
\begin{equation}\label{5}
\delta_\kappa Z^{\underline M}E_{\underline M}^{~\underline  a}=0, \quad
\delta_\kappa Z^{\underline M}E_{\underline M}^{~\underline \alpha}=
(1+\bar\Gamma)_{~\underline\beta}^{\underline\alpha}\kappa^{\underline
\beta}(\xi),
\end{equation}
where
\begin{equation}\label{61}
\bar\Gamma={1\over{6\sqrt{-g}}}\varepsilon^{mnp}
\Gamma_{mnp}\,, \qquad \bar\Gamma^2\equiv 1
\end{equation}
and hence $1+\bar\Gamma$ is a spinor projection matrix.
$\Gamma_{mnp}$ is an antisymmetric product of the target--space
gamma--matrices $(\Gamma_{\underline a})$ pulled back on to the
worldvolume, i.e
$\Gamma_m\equiv\partial_{m}Z^{\underline M}E_{\underline M}^{~\underline a}
\Gamma_{\underline a}$.

The appearance of the spinor projector in the
$\kappa$--transformations reflects
the fact that the presence of the supermembrane in the target
superspace breaks half the $2n$ supersymmetries of a
$D$--dimensional supergravity vacuum, the unbroken supersymmetries
being associated with those Grassmann coordinates
$\Theta^{ \underline\alpha}$ which can be eliminated
by $\kappa$--symmetry transformations, while remaining $n$
$\Theta^{ \a}$ are worldvolume Goldstone fermions of the
spontaneously broken supersymmetries and describe physical
fermionic modes of supermembrane fluctuations. Thus
$\kappa$--symmetry plays the same role as the worldvolume
supersymmetry of the superembedding approach (as we have
discussed in Introduction and Section 1). The exact form of the
relationship between the $\kappa$--symmetry and the
superdiffeomophisms of the superworldvolume of the supermembrame
the reader may find in \cite{s}.

An important requirement for the $\kappa$--transformations \p{5}
to be a symmetry of the membrane action \p{2} is that the
target--space supervielbeins
$E^{\underline a}(Z)$, $E^{\underline\alpha}(Z)$, superconnections
$\Omega^{~~\underline a}_{\underline b}(Z)$
and the gauge superfield
$A^{(3)}$ satisfy supergravity constraints. The essential
constraints are the torsion constraint
\begin{equation}\label{6}
T^{\underline a}=dE^{\underline a}+ E^{\underline
b}\Omega^{~~\underline a}_{\underline b} =-i\bar E_{\underline
\alpha}(\Gamma^{\underline a})^{\underline
\alpha}_{~\underline\beta}E^{\underline\beta},
\end{equation}
and the field--strength $F^{(4)}=dA^{(3)}$ constraint
\begin{equation}\label{7}
F^{(4)}={i\over 2}E^{\underline a}E^{\underline b}\bar
E_{\underline\alpha}E^{\underline\beta} (\Gamma_{\underline a
\underline b})^{\underline
\alpha}_
{~\underline\beta} +{1\over{4!}}E^{\underline a}E^{\underline b}
E^{\underline c}E^{\underline d} F_{\underline a\underline
b\underline c\underline d}.
\end{equation}
Other constraints are either conventional or can be obtained from
\p{6} and \p{7} by considering their Bianchi identities. For
example, the gauge field--strength and components of torsion are
related to each other by a constraint which in $D=11$ has the form
\begin{equation}\label{5.2}
T^{\underline\alpha}={1\over{288}} F_{\underline b_1...\underline
b_4}E^{\underline a}
\left(\Gamma_{\underline a}^{~\underline b_1...\underline b_4}-
8\delta_{\underline a}^{[\underline b_1}
\Gamma^{\underline b_2...\underline b_4]}\right)
^{\underline\alpha}_{~~\underline\beta}E^{\underline\beta}\,.
\end{equation}
In $D=4$ the first term on the right hand side of \p{5.2}
disappears because the antisymmetrized product of five
gamma--matrices is identically zero in $D=4$.

We shall assume that the supergravity constraints are imposed on
the target--superspace background also in the superembedding
description of supermembrane dynamics. Then the integrability of
the superembedding condition \p{emb} requires that the geometry
\p{e} of the $d=3$ supersurface satisfies (analogous) worldvolume supergravity
constraints, and vice versa \cite{bpstv,hs1,s}. This ensures the
consistency of the superembedding. We choose the superworldvolume
torsion constraints to be that of $N=1$, $d=3$ supergravity
\cite{d3sugra}
\begin{eqnarray}\label{tor}
T^a&=&\nabla e^a=
de^a+e^b\omega^{~a}_{b}=-i\gamma^a_{\alpha\beta}e^\alpha\wedge
e^\beta +e^b\wedge e^c\varepsilon_{cbd}\eta^{da}R(z),\\
T^\alpha&=&\nabla e^\alpha={1\over 2}e^a\wedge
e^bT^\alpha_{ba}+{i\over 2}e^a\wedge e^\beta
\gamma^\alpha_{a\beta}R(z) ,\label{toral}
\end{eqnarray}
where
$\omega^{~a}_{b}$ is a $d=3$ spin connection, $R(z)$ is an unconstrained superfield,
$\gamma^a_{\alpha\beta}$ are the $d=3$ Dirac matrices defined in
\p{Gamma1} and $\eta^{da}={\rm diag}(-,+,+)$.

\subsection{The supermembrane action in the superembedding approach}
Let us associate with the supermembrane worldvolume in
D-dimensional target superspace an $N=1$, $d=3$ supersurface
${\cal M}$ \p{z} parametrized by three bosonic $\xi^m$ and two real fermionic
(Majorana--spinor) coordinates $\eta^\mu$. It has been shown in
\cite{hrs} that the condition \p{emb} of embedding this
supersurface into an $N=1$, $D=4$ target superspace does not
contain dynamical equations of motion of the supermembrane. This
is also so for the embeddings of ${\cal M}$ into $N=1$, $D=5,11$
superspaces, which can be verified in the same way as described in
\cite{bpstv,hrs,s}.
Thus, for all these cases
\p{emb} is an off--shell constraint and one can construct an
$N=1$, $d=3$ worldvolume superfield action describing the
dynamics of the supermembrane in
$N=1$, $D=4,5,7,11$ supergravity backgrounds
\footnote{Recall that $N$ stands for the number of Majorana spinors in
$d=3$ and $D=4,11$ or Dirac spinors in $D=5,7$.}.

Recall that when we deal with the $N=1$, $d=3$ supersurface, the
$\kappa$--symmetry \p{5} is completely replaced by the worldvolume
superdiffeomorphisms only in the $N=1$, $D=4$ target superspace.
In the higher space--time dimensions a part of the
$\kappa$--transformations remains a hidden symmetry (the form of
these residual $\kappa$--transformations in the superembedding
formulation of superparticles and superstrings has been reviewed
in \cite{s}). As we have already discussed, to replace all the
$\kappa$--transformations with local worldvolume supersymmetry we
should consider $N$--extended $d=3$ supersurface (with $N=8$ in
the case of embedding into the $D=11$ target superspace), but
then the superembedding condition puts the theory on the mass
shell and the worldvolume superfield action cannot be constructed.
Since we are interested in constructing the action we choose the
supermembrane worldvolume to be the $N=1$, $d=3$ supersurface.

A worldvolume superfield form of superbrane actions can be
constructed using a generic prescription first proposed in
\cite{dghs92} for superstrings. For the supermembrane the action
of this type was constructed and analyzed in \cite{pt} (see also
\cite{hrs}). It has the following form
\begin{equation}\label{a1}
S=\int d^3\xi d^2\eta P^\alpha_{\underline a}E^{~\underline
a}_{\alpha}+ \int d^3\xi d^2\eta P^{MNP}[\tilde
A_{MNP}-(dQ)_{MNP}]\,,
\end{equation}
where the first term ensures the superembedding condition \p{emb}
as the equation of motion of the Lagrange multiplier
$P^\alpha_{\underline a}(\xi,\eta)$, and in the second term
$P^{MNP}(\xi,\eta)$ is a Lagrange multiplier, $dz^Ndz^MQ_{MN}(\xi,\eta)$ is
a superworldvolume 2--form and $\tilde A_{MNP}$ is a kind of the
pullback on to the supersurface of the following combination of
$A^{(3)}$ and $F^{(4)}=dA^{(3)}$
\begin{equation}\label{tildea}
\tilde A^{(3)}=A^{(3)}+{1\over{12}}e^a\wedge e^b\wedge e^c
\gamma_a^{\alpha\beta}E^{~\underline A}_\alpha E ^{~\underline
B}_{\beta} E^{~\underline C}_{b}E^{~\underline
D}_cF_{\underline{DCBA}}\,.
\end{equation}
$\gamma_a^{\alpha\beta}$ are $d=3$ worldvolume Dirac matrices in the
Majorana representation defined in \p{Gamma}--\p{Gamma1}. The
worldvolume form $\tilde A^{(3)}$ \p{tildea} is constructed in
such a way that it is closed ($d\tilde A^{(3)}=0$) modulo the
superembedding condition.

The action \p{a1} is classically equivalent to the supermembrane
action \p{2}. For the superworldvolume with $N>1$ Grassmann spinor
coordinates the proof was given in \cite{pt}. For the $N=1$ case
under consideration we demonstrate the equivalence in the next
subsection.

For our purposes to arrive at a worldvolume superfield action for
a supermembrane in the physical gauge the action in the form
\p{a1} is too general, since it is invariant under a huge group
of local transformations associated with the presence of the
Lagrange multipliers $P^\alpha_{\underline a}$,
$P^{MNP}$ and the auxiliary two--form field $Q_{MN}$ (see \cite{dghs92,pt}).
We should gauge fix at least a part of these symmetries. A
possible gauge fixing condition is
\begin{equation}\label{gf}
P^{MNP}={1\over{3!}}sdet(e^{~A}_L)~e^{~M}_ae^{~N}_\alpha
e^{~P}_\beta\gamma^{a\alpha\beta},
\end{equation}
where $e^{~M}_A(z)$ is inverse of the worldvolume supervielbein
matrix \p{e}. Substituting \p{gf} into \p{a1} we reduce the
action to the following form
\begin{equation}\label{a2}
S=\int d^3\xi d^2\eta P^\alpha_{\underline a}E^{~\underline
a}_{\alpha}+ {1\over{3!}}\int d^3\xi
d^2\eta~sdet{~e}~\gamma^{a\alpha\beta}A_{\alpha\beta a}\,,
\end{equation}
where
\begin{equation}\label{pulla}
A_{\alpha\beta a}=E^{~\underline C}_\alpha E^{~\underline
B}_\beta E^{~\underline A}_a A_{\underline{ABC}}=3E^{~\underline
\alpha}_\alpha E^{~\underline \beta}_\beta E^{~\underline \gamma}_a
A_{\underline\alpha\underline\beta\underline\gamma}+E^{~\underline
\alpha}_\alpha E^{~\underline \beta}_\beta E^{~\underline a}_a
A_{\underline\alpha\underline\beta\underline a}+ \cdots
\end{equation}
The dots in \p{pulla} stand for the terms containing the
$E^{~\underline a}_\alpha(z)$ components of the pullback of the
target--space supervielbein $E^{\underline a}$ \p{pull}. These
terms contribute to the first term of \p{a2} (which simply results
in the redefinition of $P^\alpha_{\underline a}$) and hence can be
ignored.

The action \p{a2} is simpler and looks much more attractive than
\p{a1}. Its second term (which actually produces the dynamical equations of motion
of the supermembrane) does not contain Lagrange multipliers and
resembles the Wess--Zumino term of the action \p{2}. Indeed, upon
integrating over the Grassmann--odd coordinates and eliminating
the auxiliary fields (by the use of the superembedding condition
incorporated in the first term), the second term of \p{a2}
produces both, the Nambu--Goto and the Wess--Zumino part of \p{2}.
We may also note that because of dimensional reasons the choice
of the co--dimension two component of the pullback of
$A^{(3)}$ for the construction of the action is unique.
For superstrings a similar form of the superembedding action was
proposed in \cite{to}.

By construction the action \p{a2} is manifestly invariant under
the worldvolume and target space superdiffeomorphisms and local
$SO(1,2)$ rotations in the superworldvolume tangent superspace.
In addition it is also invariant under the following super--Weyl
transformations of the components of the worldvolume supervielbein
\begin{equation}\label{superW}
{e'}^a=W^2(z)e^a, \quad
{e'}^\alpha=W(z)e^\alpha-ie^a\gamma_a^{\alpha\beta}{\cal D}_\beta
W,
\end{equation}
and its inverse
\begin{equation}\label{super-W}
{e'}^{~M}_a=W^{-2}{e}^{~M}_a-W^{-3}{\cal D}_\alpha W
\gamma_a^{\alpha\beta}e^{~M}_\beta,
\quad {e'}^{~M}_\alpha=W^{-1}{e}^{~M}_\alpha,
\end{equation}
where ${\cal D}_\alpha=e^{~M}_\alpha\partial_M$. Note that the
super--Weyl transformations \p{superW} leave intact the torsion
constraint \p{tor} (i.e.
$T^a_{\alpha\beta}=-2i\gamma^a_{\alpha\beta}$).

The invariance of \p{a2} under \p{superW}, \p{super-W} can be
easily verified using the following form of the superdeterminant
\begin{equation}\label{sdet}
{sdet}~e^{~A}_M={sdet}^{-1}~e^{~M}_A=det^{-1}[e^{~m}_a-
e^{~\mu}_a(e^{~\alpha}_\mu)^{-1}e^{~m}_\alpha]~det~e^{~\mu}_\alpha,
\end{equation}
(where $(e^{~\alpha}_\mu)^{-1}$ is inverse of $e^{~\mu}_\alpha$)
from which it follows that under \p{superW}, \p{super-W} rescales
as
\begin{equation}\label{sdW}
{sdet}~e'=W^4{sdet}~e.
\end{equation}
The super--Weyl variation of the Lagrangian density
$\gamma^{a\alpha\beta}A_{\alpha\beta a}$ of \p{a2} is
\begin{equation}\label{AW}
\gamma^{a\alpha\beta}A'_{\alpha\beta a}=W^{-4}\gamma^{a\alpha\beta}A_{\alpha\beta a}
+W^{-5}{\cal
D}_{\delta}W\gamma_a^{\delta\gamma}\gamma^{a\beta\alpha}E^{~\underline\alpha}_\alpha
E^{~\underline\beta}_\beta E^{~\underline\gamma}_\gamma
A_{\underline\alpha\underline\beta\underline\gamma}+ \cdots.
\end{equation}
The second term in \p{AW} vanishes due to the $d=3$ gamma--matrix
cyclic identity
\begin{equation}\label{cycle}
\gamma_a^{\delta\gamma}\gamma^{a\beta\alpha}+\gamma_a^{\delta\beta}\gamma^{a\alpha\gamma}
+\gamma_a^{\delta\alpha}\gamma^{a\gamma\beta}=0,
\end{equation}
and dots stand for a term proportional to
$E^{~\underline a}_\alpha$ which can be canceled by
an appropriate variation of the Lagrange multiplier
$P^{\alpha}_{\underline a}$ in \p{a2}.

We have thus demonstrated that the supermembrane action \p{a2} is
invariant under the super--Weyl transformations \p{superW},
\p{super-W} of the superworldvolume. Note that conventional
(super)membrane actions do not have such symmetry.

The local $SO(1,2)$ rotations and the super--Weyl transformations
can be used to put the components $e^{~\mu}_\alpha$ of the
inverse supervielbein matrix $e^{~M}_A$ to be equal to the unit
matrix
\begin{equation}\label{e-1re}
e^{~M}_A=\left(
\begin{array}{cc}
e^{~m}_a &  e^{~\mu}_a\\
e^{~m}_\alpha &  \delta^\mu_\alpha
\end{array}
\right)\, ,
\end{equation}
then the superdeterminant \p{sdet} reduces to
\begin{equation}\label{sdetr}
{sdet}~e^{~A}_M=det^{-1}[e^{~m}_a- e^{~\alpha}_a e^{~m}_\alpha].
\end{equation}
We shall further work in the gauge \p{e-1re}, \p{sdetr}.

\subsection{Relationship with the conventional formulation}
To establish the relationship we should consider the second term
of \p{a2}, since the first term only serves for producing the
superembedding constraint which relates the superworldvolume
geometry with that of the target superspace, in other words, which
ensures the components of the  worldvolume supergravity multiplet
to be pure auxiliary fields. In this sense the local worldvolume
supersymmetric action \p{a2} is an example of how a `no--go'
theorem \cite{bst2} of the (non)existence of local worldvolume
supersymmetric extensions of the Dirac membrane action (and, in
particular, of its Howe--Tucker form \cite{ht}) can be overcome.

It is well known that integration over the Grassmann--odd
variables is equivalent to differentiation. So (taking into
account the superembedding condition \p{emb} and the
superworldvolume constraints \p{tor} and \p{toral}) we rewrite
$$
S_A={1\over{3!}}\int d^3\xi
d^2\eta~sdet{~e}~\gamma^{a\alpha\beta}A_{\alpha\beta a}={1\over{2i
3!}}\int d^3\xi
\partial^\gamma\partial_\gamma~(sdet{~e}~\gamma^{a\alpha\beta}A_{\alpha\beta
a})
$$
\begin{equation}\label{a2A}
={1\over{2i 3!}}\int d^3\xi
~sdet{~e}~\nabla^\gamma\nabla_\gamma\gamma^{a\alpha\beta}A_{\alpha\beta
a}~|_{\eta=0}
\end{equation}
Upon some manipulations with the use of the supergravity
constraints \p{6}--\p{5.2} and \p{tor}, and the superembedding
condition \p{emb}, one finds that the second--order covariant
derivative in \p{a2A} is
$$
S_A=-{1\over{3!}}\int d^3\xi~sdet~e~[{i\over
2}\varepsilon^{abc}\gamma_c^{\alpha\beta}
E^{\underline\alpha}_\alpha E^{\underline\beta}_\beta
E^{\underline b}_bE^{\underline a}_a
F_{\underline\alpha\underline\beta\underline a\underline
b}+{1\over 2}\varepsilon^{abc}A_{cba}]|_{\eta=0}
$$
\begin{equation}\label{a2A1}
=-\int d^3\xi~sdet~e~[{1\over{2\cdot 3!}}
\varepsilon^{abc}\gamma_c^{\alpha\beta}
E^{\underline\alpha}_\alpha E^{\underline\beta}_\beta
E^{\underline b}_bE^{\underline a}_a (\Gamma_{\underline
a\underline
b})_{\underline\alpha\underline\beta}]|_{\eta=0}-{1\over{2\cdot
3!}}\int d^3\xi~sdet~e~ \varepsilon^{abc}A_{cba} |_{\eta=0}\,.
\end{equation}

To reduce the action \p{a2A1} to the Green--Schwarz action we
choose a Wess--Zumino gauge such that
\begin{equation}\label{wz}
e^{~\mu}_a|_{\eta=0}=0.
\end{equation}
Then the leading component of the superdeterminant
\p{sdetr} reduces to $det^{-1}e_a^{~m}(\xi)$ and one can easily
see that the second term of \p{a2A1} coincides with the
Wess--Zumino term of \p{2}.

To show that the first term of \p{a2A1} is equivalent to the
Nambu--Goto term we use the harmonic technics of the
superembedding approach
\cite{bpstv,hrs,s}. In the case of the supermembrane in $N=1$, $D=4,5,7$ and
11, with our choice \p{tor} of the worldvolume supergeometry
constraints, the superembedding condition \p{emb} allows us to
choose the pullback of the target--space supervielbein \p{E} to be
\begin{equation}\label{Eap}
E^{\underline a}=e^a E_a^{~\underline a}=(1+h^{\dot q}h^{\dot q})
e^au^{~\underline a}_a
\end{equation}
\begin{equation}\label{Ealp}
E^{\underline\alpha}=e^\alpha
E_\alpha^{~\underline\alpha}+e^aE_a^{~\underline\alpha}=e^\alpha
v^{~\underline\alpha}_{\alpha p}~n^p + h^{\dot q}(z)~{\tilde
v}^{~\underline\alpha}_{\alpha\dot q} +e^aE_a^{~\underline\alpha},
\end{equation}
 where $u^{~\underline a}_a(z)$, and
$v^{~\underline\alpha}_{\alpha p}(z)$ and ${\tilde
v}^{~\underline\alpha}_{\alpha\dot q}(z)$ are, respectively,
vector and spinor harmonics parametrizing the coset space
$SO(1,D-1)/[SO(1,2)\times SO(D-3)]$ with indices
$(a,\alpha)$ being associated with the vector and spinor representation of
$SO(1,2)$ and the indices $(p,\dot q)$ corresponding to (in general
different) ($D-2)$--dimensional spinor representations of
$SO(D-3)$.
$n^p$ is a constant unit--norm spinor ($n^pn^p=1$) and $h^{\dot
q}(z)$ is an unconstrained worldvolume superfield.

The harmonics have the following properties (see \cite{bpstv,s}
for a review)
\begin{equation}\label{uv}
u^{~\underline a}_au^{~\underline
b}_b\eta_{\underline{ab}}=\eta_{ab}, \quad
v^{~\underline\alpha}_{\alpha p}~v^{~\underline\beta}_{\beta
q}~C_{\underline\alpha\underline\beta}=\epsilon_{\alpha\beta}\delta_{pq},
\quad
{\tilde v}^{~\underline\alpha}_{\alpha \dot p}~{\tilde
v}^{~\underline\beta}_{\beta
\dot q}~C_{\underline\alpha\underline\beta}=\epsilon_{\alpha\beta}\delta_{\dot p\dot
q},
\quad
v^{~\underline\alpha}_{\alpha p}~{\tilde
v}^{~\underline\beta}_{\alpha
\dot q}~C_{\underline\alpha\underline\beta}=0,
\end{equation}
\begin{equation}\label{ugv}
(v^{~\underline\alpha}_{\alpha p}~v^{~\underline\beta}_{\beta
p}+{\tilde v}^{~\underline\alpha}_{\alpha
\dot q}~{\tilde v}^{~\underline\beta}_{\beta \dot q})\gamma^{a\alpha\beta}
=\Gamma_{\underline
b}^{\underline\alpha\underline\beta}~u^{{\underline b}\,\,a}.
\end{equation}

Using \p{wz}, \p{Eap} and \p{uv} one finds that the induced
metric is related to the bosonic vielbein matrix $e^{~a}_m$,
inverse of
$e^{~m}_a$, as follows
\begin{equation}\label{ind2}
g_{mn}(\xi)=E^{~\underline a}_mE^{\underline
b}_n~\eta_{\underline{ab}}|_{\eta=0}=(1+h^{\dot q}h^{\dot
q})^2~e^{~a}_me_{an}|_{\eta=0},
\end{equation}
and, hence
\begin{equation}\label{sdeto}
sdet ~e|_{\eta=0}=det ~e^{~a}_m(\xi)={1\over{(1+h^{\dot q}h^{\dot
q})^3}}\sqrt{-det~g_{mn}}.
\end{equation}
Finally, using the expressions of $E^{~\underline a}_a$ and
$E^{~\underline \alpha}_\alpha$ in terms of the harmonics and of the
superfield $h^{\dot q}(z)$ \p{Eap}, \p{Ealp} and the relations
\p{uv} and \p{ugv}, as well as the
gamma--matrix identity
$(\gamma^{[a}\gamma^b\gamma^{c]})^\alpha_{~\beta}=\delta^\alpha_\beta\varepsilon^{abc}$,
one reduces the action \p{a2A1} to
\begin{equation}\label{a2A2}
S=\int d^3\xi \sqrt{-det~g_{mn}}~{{1-h^{\dot q}h^{\dot
q}}\over{1+h^{\dot q}h^{\dot q}}}-{1\over 2}
\int_{{\cal M}_3}d^3\xi
\varepsilon^{mnp}\partial_mZ^{\underline L}\partial_nZ^{\underline M}
\partial_pZ^{\underline N}A_{\underline{NML}}(Z).
\end{equation}
Varying eq. \p{a2A2} with respect to $h^{\dot q}$ we find that its
algebraic equation of motion implies $h^{\dot q}=0$~\footnote{This
is so, if we assume that the induced metric is non--degenerate.
Otherwise we would get a tensionless (null) supermembrane.}, and
thus \p{a2A2} reduces to the conventional supermembrane action
\p{2}.

Note that in the $D=4$ target--superspace the group $SO(D-3)$
gets trivialized so in this case we have only one scalar
superfield
$h(z)$, and the supermembrane action \p{a2A2} coincides with the
one constructed in \cite{hrs}.

\subsection{$D=4$ supermembrane in the physical gauge and
spontaneous supersymmetry breaking}
 We now proceed to the consideration of the dynamics of a
supermembrane propagating in an $N=1$, $D=4$ flat target
superspace in the physical gauge \p{static}.

In the flat target superspace of dimension $D=4,5,7$ and 11 the
three--form field is
\begin{equation}\label{Aflat}
A^{(3)}=i\bar\Theta\Gamma_{\underline{ab}}d\Theta({\cal
E}^{\underline a}{\cal E}^{\underline b}-i{\cal E}^{\underline a}
\bar\Theta\Gamma^{\underline{b}}d\Theta-
{1\over
3}\bar\Theta\Gamma^{\underline{a}}d\Theta\bar\Theta\Gamma^{\underline{b}}d\Theta),
\end{equation}
where
\begin{equation}\label{calE}
{\cal E}^{\underline a}={dX}^{\underline
a}-id\bar\Theta\Gamma^{\underline{a}}\Theta.
\end{equation}

The action \p{a2} takes the form
$$
S=\int d^3\xi d^2\eta P^\alpha_{\underline a}{\cal E}^{~\underline
a}_{\alpha} +{1\over{3!}}\int d^3\xi
d^2\eta~sdet{~e}~\gamma^{a\alpha\beta}[i\bar\Theta\Gamma_{\underline{ab}}{\cal
D}_\alpha
\Theta(i
\bar\Theta\Gamma^{\underline{a}}{\cal D}_\beta\Theta {\cal E}^{\underline b}_a
$$
\begin{equation}\label{a2flat}
+{2\over 3}\bar\Theta\Gamma^{\underline{a}}{\cal D}_\beta
\Theta\bar\Theta\Gamma^{\underline{b}}{\cal D}_a\Theta)+
{1\over 3}i\bar\Theta\Gamma_{\underline{ab}}{\cal D}_a
\Theta\bar\Theta\Gamma^{\underline{a}}{\cal D}_\alpha
\Theta\bar\Theta\Gamma^{\underline{b}}{\cal D}_\beta\Theta],
\end{equation}
where the components of \p{Aflat} containing
${\cal E}^{~\underline a}_\alpha$
have been included into the first term, and
${\cal D}_a=e^{~M}_a\partial_M$. (Recall that ${\cal E}^{~\underline a}_\alpha=0$
is the superembedding condition.)

Using the cyclic identity for gamma--matrices in $D=4,5,7$ and
$11$
\begin{equation}\label{4511}
\Gamma_{\underline a(\underline\alpha\underline\beta}
\Gamma^{\underline{ab}}_{\underline\gamma\underline\delta)}
 =0
\end{equation}
(where () denotes the symmetrization of the spinor indices) we
can reduce the second `Wess--Zumino' term of the action
\p{a2flat} to
\begin{equation}\label{a2flatr}
S_A={i\over{3!}}\int d^3\xi
d^2\eta~sdet{~e}~\gamma^{a\alpha\beta}\bar\Theta\Gamma_{\underline{ab}}{\cal
D}_\alpha
\Theta(i
\bar\Theta\Gamma^{\underline{a}}{\cal D}_\beta\Theta {\cal E}^{\underline
b}_a+\bar\Theta\Gamma^{\underline{a}}{\cal D}_\beta
\Theta\bar\Theta\Gamma^{\underline{b}}{\cal D}_a\Theta).
\end{equation}

In $D=4$ the supermembrane action can be further simplified due to
the one more cyclic gamma--matrix identity similar to \p{cycle}.
In particular, we find that
\begin{equation}\label{d4i}
\bar\Theta\Gamma_{\underline a\underline
b}d\Theta\bar\Theta\Gamma^{\underline b}d\Theta=-{1\over
2}(\bar\Theta\Theta)d\bar\Theta\Gamma_{\underline a}d\Theta,
\end{equation}
and, hence,
\begin{equation}\label{Aflatd4}
A^{(3)}=i\bar\Theta\Gamma_{\underline{ab}}d\Theta{\cal
E}^{\underline a}{\cal E}^{\underline b}+{i\over
2}(\bar\Theta\Theta)d\bar\Theta\Gamma_{\underline{a}}d\Theta{\cal
E}^{\underline a}.
\end{equation}
Substituting the $A_{\alpha\beta a}$ component of the
superworldvolume pullback of \p{Aflatd4} into the action \p{a2}
we get the $D=4$ supermembrane action in the form
\begin{equation}\label{a2d4}
S=\int d^3\xi d^2\eta P^\alpha_{\underline a}{\cal E}^{~\underline
a}_{\alpha}+{i\over{2\cdot 3!}}\int d^3\xi
d^2\eta~sdet{~e}~\gamma^{a\alpha\beta}~(\bar\Theta\Theta) ~{\cal
D}_\alpha
\bar\Theta\Gamma_{\underline{a}}{\cal D}_\beta\Theta{\cal E}^{~\underline
a}_a.
\end{equation}

The form \p{a2d4} of the supermembrane action and of the
superdeterminant \p{sdetr} prompt us that it is invariant under
the following variation of supervielbein components \p{e-1re}
\begin{equation}\label{vare}
{e'}^{~\mu}_a={e}^{~\mu}_a+f^{~\mu}_a(z), \quad
{e'}^{~m}_a=e^{~m}_a+f^{~\alpha}_ae^{~m}_\alpha,
\end{equation}
accompanied by an appropriate variation of the Lagrange multiplier
$P_{\underline a}^\alpha$ \footnote{Of course the variation \p{vare} changes conventional
worldvolume supergravity constraints in \p{tor} and \p{toral}, but
the esseintial constraint
$T_{\alpha\beta}^a=-2i\gamma^a_{\alpha\beta}$ remains unchanged.}. This allows us to
put
$e^{~\mu}_a=0$, and thus reduce
\p{e-1re} and \p{sdetr} to
\begin{equation}\label{e-1re1}
e^{~M}_A=\left(
\begin{array}{cc}
e^{~m}_a &  0\\
e^{~m}_\alpha &  \delta^\mu_\alpha
\end{array}
\right)\, , \quad {sdet}~e^{~A}_M=det^{-1}(e^{~m}_a).
\end{equation}

We can also notice that the integrability of the superembedding
condition
\begin{equation}\label{embflat}
{\cal E}^{~\underline a}_\alpha={\cal D}_\alpha X^{\underline
a}-i{\cal D}_\alpha\bar\Theta\Gamma^{\underline a}\Theta=0
\end{equation}
requires that
\begin{equation}\label{integ}
\gamma^a_{\alpha\beta}{\cal E}^{~\underline a}_a
=\gamma^a_{\alpha\beta}({\cal D}_a X^{\underline a}
-i{\cal D}_a\bar\Theta\Gamma^{\underline a}\Theta)={\cal
D}_\alpha\bar\Theta\Gamma^{\underline a}{\cal D}_\beta\Theta.
\end{equation}
Eq. \p{integ} is obtained from \p{embflat} by hitting its right
hand side with $\nabla_\alpha={\cal D}_\beta+\omega_\beta$,
symmetrizing the result with respect to indices $\alpha,\beta$
and taking into account that due to the torsion constraint \p{tor}
\begin{equation}\label{anti}
\{\nabla_\alpha,\nabla_\beta\}=2i\gamma^a_{\alpha\beta}\nabla_a
-T^\gamma_{\alpha\beta}\nabla_\gamma+R_{\alpha\beta},
\end{equation}
where $R_{\alpha\beta}^{AB}(z)$ are components of the
superworldvolume curvature.

Using eq. \p{integ} we can rewrite the action \p{a2d4} in even
simpler form
\begin{equation}\label{a2d4p}
S=\int d^3\xi d^2\eta P^\alpha_{\underline a}{\cal E}^{~\underline
a}_{\alpha}-{i\over{3!}}\int d^3\xi
d^2\eta~det^{-1}(e^{~m}_a)~(\bar\Theta\Theta)~{\cal
E}^{~\underline a}_a{\cal E}^{~\underline b a}\eta_{\underline
a\underline b},
\end{equation}
$\eta_{\underline a\underline b}={\rm diag}~ (-,+,+,+)$.

 Note that in the form \p{a2d4p} the action resembles the
 Howe--Tucker--Polyakov term of the supermembrane action.

We now use the worldvolume superdiffeomorphisms to impose the
physical gauge \p{static}. To this end we choose the following
`$d=3$ adapted' Majorana representation of the $D=4$ Dirac
matrices
$\Gamma^{\underline a}=(\Gamma^a,\Gamma^3)$ $(a=0,1,2)$
\begin{equation}\label{Gamma4}
\Gamma^{a}_{\underline\alpha\underline\beta}= \left(
\begin{array}{cc}
\gamma^a_{\alpha\beta} &  0\\
0 &(\gamma^a)^{\alpha\beta}
\end{array}
\right)\,, \quad
\Gamma^{3}_{\underline\alpha\underline\beta}=
\left(
\begin{array}{cc}
0  &  \delta^\alpha_\beta\\
\delta^\beta_\alpha &  0
\end{array}
\right)\, , \quad C_{\underline\alpha\underline\beta}=\left(
\begin{array}{cc}
\epsilon_{\alpha\beta}  &  0\\
0   &   -\epsilon^{\alpha\beta}
\end{array}
\right) \,,
\end{equation}
where $\gamma^a_{\alpha\beta}$ are the same as defined in
\p{Gamma1}.

With respect to \p{Gamma4} the Majorana spinor
$\Theta^{\underline\alpha}$ ($\alpha=1,\cdots, 4$) splits as
\begin{equation}\label{ss}
\Theta^{\underline\alpha}(z)=
\left(
\begin{array}{c}
\theta^\alpha \\
\Psi_\beta
\end{array}
\right)\,.
\end{equation}

In the physical gauge we identify target superspace coordinates
$X^a$ and $\theta^\alpha$ with superworldvolume coordinates
\begin{equation}\label{static2}
X^a=\xi^a, \quad, \theta^\alpha=\eta^\alpha\,.
\end{equation}
Upon this identification there is no distinction between
worldvolume indices $(m,\mu)$ and tangent superspace indices
$(a,\alpha)$.
The remaining superfields $X^3(z)$ and $\Psi_\alpha(z)$ are
Goldstones of spontaneously broken space--time translations in the
direction transverse to the membrane and of two supersymmetry
transformations. In the gauge \p{static2} the superembedding
condition splits into the part parallel to the membrane (remember
that the worldvolume supervielbein matrix has the form \p{e-1re1})
\begin{equation}\label{para}
e^{~m}_\alpha-i\gamma^m_{\alpha\beta}\eta^\beta-i{\cal
D}_\alpha\bar\Psi\gamma^m\Psi=0
\end{equation}
and the transverse part
\begin{equation}\label{tran}
{\cal D}_\alpha \Phi(z)=2i\Psi_\alpha(z),
\end{equation}
where
\begin{equation}\label{X3}
\Phi=X^3+i\eta^\alpha\Psi_\alpha, \quad {\cal
D}_\alpha=\partial_\alpha+e^{~m}_\alpha\partial_m.
\end{equation}

The integrability condition \p{integ} splits as follows. The
parallel part is
\begin{equation}\label{pari}
e^{~m}_a-i{\cal D}_a\Psi\gamma^m\Psi=\delta^m_a-{1\over
2}\gamma_a^{\alpha\beta}{\cal D}_\alpha\bar\Psi\gamma^m{\cal
D}_\beta\Psi
\end{equation}
and the transverse part is
\begin{equation}\label{transi}
{\cal
D}_a\Phi=-\gamma^{\alpha\beta}_a{\cal D}_\alpha\Psi_\beta, \quad
{\cal D}_a=e^{~m}_a\partial_m.
\end{equation}

 Now eqs. \p{para} and \p{pari} can be rewritten in the form
$$
e^{~n}_\alpha(\delta^m_n-i\partial_n\bar\Psi\gamma^m\Psi)=
i\gamma^m_{\alpha\beta}\eta^\beta+i\partial_\alpha\bar\Psi\gamma^m\Psi.
$$
\begin{equation}\label{para1}
e^{~n}_a(\delta_n^m-i\partial_n\bar\Psi\gamma^m\Psi)=\delta^m_a-{1\over
2}\gamma_a^{\alpha\beta}{\cal D}_\alpha\bar\Psi\gamma^m{\cal
D}_\beta\Psi,
\end{equation}
The inverse of the matrix
$M_n^{~m}=\delta^m_n-i\partial_n\Psi\gamma^m\Psi$
is
$$
(M^{-1})^{~m}_n=\delta^m_n+i\partial_n\bar\Psi\gamma^m\Psi-\partial_n\bar\Psi\gamma^b\Psi
\partial_b\bar\Psi\gamma^m\Psi.
$$
Then from \p{para1} we get the expression for the worldvolume
supervielbein components $e^{~n}_A(z)$ in terms of the Goldstone
fermion $\Psi_\alpha(z)$
$$
e^{~n}_\alpha=i\gamma^n_{\alpha\beta}\eta^\beta+iD_\alpha\bar\Psi\gamma^n\Psi-
D_\alpha\bar\Psi\gamma^b\Psi
\partial_b\bar\Psi\gamma^n\Psi
$$
\begin{equation}\label{epsi}
=i\gamma^n_{\alpha\beta}\eta^\beta+iD_\alpha\bar\Psi\gamma^b\Psi
(\delta_b^n+i\partial_b\bar\Psi\gamma^n\Psi),
\end{equation}
\begin{equation}\label{ea}
e^{~n}_a=(\delta_a^b-{1\over 2}\gamma_a^{\alpha\beta}{\cal
D}_\alpha\bar\Psi\gamma^b{\cal D}_\beta\Psi)
(\delta_b^n+i\partial_b\bar\Psi\gamma^n\Psi-\partial_b\bar\Psi\gamma^m\Psi
\partial_m\bar\Psi\gamma^n\Psi),
\end{equation}
where
\begin{equation}\label{Da}
D_\alpha={\partial\over{\partial\eta^\alpha}}+i\eta^\beta\gamma^a_{\beta\alpha}
{\partial\over{\partial\xi^a}}, \quad \{D_\alpha,D_\beta\}=2i\gamma^a_{\alpha\beta}
{\partial\over{\partial\xi^a}}
\end{equation}
are covariant derivatives in a flat $N=1$, $d=3$ superspace
\footnote{Up to a normalization our definition of \p{epsi} and ${\cal D}_\alpha$ \p{X3}
is the same as in \cite{ik1}, and the definition of \p{ea} and
${\cal D}_a$ \p{transi} is related to that in \cite{ik1} by the linear transformation
with the matrix $(\delta_a^b-{1\over 2}\gamma_a^{\alpha\beta}{\cal
D}_\alpha\bar\Psi\gamma^b{\cal D}_\beta\Psi)$.}.

To complete the list of expressions for the $e^{~M}_A$ components
we also give the equation relating $e^{~m}_a$ and
$e^{~m}_\alpha$
\begin{equation}\label{vs}
e^{~m}_a={i\over 2}\gamma_a^{\alpha\beta}{\cal D}_\beta
e^{~m}_\alpha,
\end{equation}
which can be easily obtained by taking the ${\cal
D}_\beta$--derivative of \p{para}, symmetrizing the result with
respect to $\alpha,\beta$ and comparing it with eq. \p{pari}.

We have thus expressed all components of the worldvolume
supervielbein \p{e-1re1} in terms of the Goldstone superfield,
which implies that the geometry (supergravity) in the
superworldvolume is indeed induced by its embedding into the
target superspace.

Consider now the transverse part \p{tran} of the superembedding
condition. In view of \p{epsi} it can be presented in the
following form
\begin{equation}\label{tran1}
\Psi_\alpha=-{i\over 2}D_\alpha\Phi+{1\over 2}D_\alpha\bar\Psi\gamma^b\Psi
(\delta_b^n+i\partial_b\bar\Psi\gamma^n\Psi)\partial_n\Phi.
\end{equation}
This is an implicit expression of the Goldstone fermion
$\Psi_\alpha(\xi,\eta)$ in terms of the independent Goldstone boson
superfield $\Phi(\xi,\eta)$. Eq. \p{tran1} is exactly solvable.
However, the solution looks rather cumbersome and we only present
its general structure
\begin{equation}\label{sol}
\Psi_\alpha=iF_\alpha^{~\beta}D_\beta\Phi+\tilde F_\alpha ~D^\beta\Phi
D_\beta\Phi,
\end{equation}
where $F_\alpha^{~\beta}=F_1\delta_\alpha^{~\beta}
+F_2(\partial_a\Phi\gamma^a)_\alpha^{~\beta}
$ and
$\tilde F_\alpha$ are known though complicated functions of
$\partial_m\Phi$ and $D_\alpha\Phi$.

To find a form of the supermembrane action \p{a2d4p} in the
physical gauge we should calculate ${\cal E}^{~\underline
a}_a{\cal E}^{~\underline b a}\eta_{\underline a\underline b}$
and $det(e_a^{~m})$ using the expressions \p{tran}, \p{pari},
\p{transi} and
\p{ea}. To this end it is convenient to rewrite the matrix
\p{pari} in the following form
\begin{equation}\label{pari1}
L_a^{~m}\equiv(\delta_a^m-{1\over 2}\gamma_a^{\alpha\beta}{\cal
D}_\alpha\bar\Psi\gamma^m{\cal D}_\beta\Psi)=\delta^m_a+{1\over
4}\delta^m_a[({\cal D}_\alpha\Psi^\alpha)^2+{\cal D}^c\Phi{\cal
D}_c\Phi]-{1\over 2}{\cal D}_a\Phi{\cal D}^m\Phi+{1\over
2}\varepsilon_{a}^{~mc}{\cal D}_c\Phi({\cal D}_\alpha\Psi^\alpha).
\end{equation}
Then one finds that
\begin{equation}\label{EE}
{\cal E}^{~\underline
a}_a{\cal E}^{~\underline b a}\eta_{\underline a\underline b}=
3[1+{1\over 4}({\cal D}_\alpha\Psi^\alpha)^2+{1\over 4}{\cal
D}^a\Phi{\cal D}_a\Phi]^2-{3\over 4}({\cal
D}_\alpha\Psi^\alpha)^2{\cal D}^a\Phi{\cal D}_a\Phi,
\end{equation}
\begin{eqnarray}\label{dete}
&&det(e^{~m}_a)=det(L_a^{~m})~
det^{-1}(\delta^b_m-i\partial_m\bar\Psi\gamma^b\Psi)\nonumber\\
&=&{1\over 3}{\cal E}^{~\underline a}_a{\cal E}^{~\underline b
a}\eta_{\underline a\underline b}\left[1+{1\over 4}({\cal
D}_\alpha\Psi^\alpha)^2-{1\over 4}{\cal D}^c\Phi{\cal
D}_c\Phi\right]
det^{-1}(\delta^b_m-i\partial_m\bar\Psi\gamma^b\Psi),
\end{eqnarray}
and
\begin{equation}\label{det}
det(\delta^b_m-i\partial_m\bar\Psi\gamma^b\Psi)=
1-i\partial_a\bar\Psi\gamma^a\Psi +{1\over 2}
\epsilon^{abc}\partial_a\bar\Psi\gamma_b\partial_c\Psi(\Psi)^2,
\end{equation}
(where $(\Psi)^2=\Psi^\alpha\Psi_\alpha$),  and the action
\p{a2d4p} takes the form
\begin{equation}\label{aphys}
S=-{i\over 2}\int d^3\xi
d^2\eta~(\eta^2-\Psi^2)~{{det(\delta^b_m-i\partial_m\bar\Psi\gamma^b\Psi)}
\over{1+{1\over
4}({\cal D}_\alpha\Psi^\alpha)^2-{1\over 4}{\cal D}^c\Phi{\cal
D}_c\Phi}},
\end{equation}
Upon some algebraic manipulations with the use of eqs. \p{epsi}
the denominator of
\p{aphys} can be represented as follows

$
1+{1\over 4}({\cal D}_\alpha\Psi^\alpha)^2-{1\over 4}{\cal
D}^c\Phi{\cal D}_c\Phi=
$
\begin{equation}\label{deno}
=1+{1\over 4}\left[(D_\alpha\Psi^\alpha)^2-({\bar
D}\gamma_a\Psi)({\bar
D}\gamma^a\Psi)\right]det(\delta^b_m-i\partial_m\bar\Psi\gamma^b\Psi)\,.
\end{equation}
Note also that
\begin{equation}\label{dpsi}
(D_\alpha\Psi^\alpha)^2-({\bar D}\gamma_a\Psi)({\bar
D}\gamma^a\Psi)=2D_\alpha\Psi_\beta
D^\alpha\Psi^\beta=-D^2\Psi^2+2\Psi^\alpha D^2\Psi_\alpha, \quad
D^2\equiv D^\alpha
D_\alpha.
\end{equation}

As we have already observed in the case of the superparticles
(see eqs. \p{d2gauge} and \p{Df}), the manifest worldvolume and
target--space supersymmetry of the original action \p{a2}, and
dimensional reasons requires the fractional factor in the action
\p{aphys} to be of the form
\begin{equation}\label{frac}
{\cal L}\equiv {{det(\delta^b_m-i\partial_m\bar\Psi\gamma^b\Psi)}
\over{1+{1\over
4}({\cal D}_\alpha\Psi^\alpha)^2-{1\over 4}{\cal D}^c\Phi{\cal
D}_c\Phi}}=D^{\alpha}D_\alpha~{\Psi^2\over{Y(\xi,\eta)}}+
\partial_a(\Psi^2Y^a)+1,
\end{equation}
where $Y(\xi,\eta)$ and $Y^a(\xi,\eta)$ are  superfields,  and 1
reflects the fact that the energy density of the ``ground'' state
($\Phi=const,~\Psi=0$) of the supermembrane is normalized to be
one (in tension unites), which is in accordance with the value of
the energy density of a (non--fluctuating) supermembrane ground
state in the Green--Schwarz formulation \p{2}.

Note that because of its form the vector derivative term of
\p{frac} appears in the action \p{aphys} only as a total
derivative, and, therefore, can be omitted.

We now show how to determine the form of the superfield
$Y(\xi,\eta)$. To this end we take the part ${i\over 2}\int d^3\xi
d^2\eta~\Psi^2{\cal L}$ of the action \p{aphys}. Because of the
relations \p{deno}--\p{dpsi}
\begin{equation}\label{L1}
{i\over 2}\int d^3\xi d^2\eta~\Psi^2{\cal L}= {i\over 2}\int
d^3\xi d^2\eta~{\Psi^2\over{1-{1\over 4}D^2\Psi^2}}.
\end{equation}
On the other hand from \p{frac} we get (up to a total derivative)
that
\begin{equation}\label{L2}
{i\over 2}\int d^3\xi d^2\eta~\Psi^2{\cal L}= {i\over 2}\int
d^3\xi
d^2\eta~\left[(D^2\Psi^2){\Psi^2\over{Y(\xi,\eta)}}+\Psi^2\right].
\end{equation}
Comparing \p{L1} with \p{L2} we find that
$$
Y=4~(1-{1\over 4}D^2\Psi^2),
$$
and
\begin{equation}\label{frac1}
{\cal L}=1+{1\over 4}D^2{\Psi^2\over{1-{1\over 4}D^2\Psi^2}}.
\end{equation}
Substituting \p{frac1} into \p{aphys} we get the following
$N=1$, $D=3$ Goldstone superfield action
\begin{equation}\label{aphys1}
S=i\int d^3\xi d^2\eta~{\Psi^2\over{1-{1\over 4}D^2\Psi^2}}+\int
d^3\xi\cdot 1\, ,
\end{equation}
where the Goldstone fermion $\Psi_\alpha$ depends on the
Goldstone scalar $\Phi$
\p{tran}, \p{sol}.

Finally we present the action in terms of the independent
Goldstone scalar superfield $\Phi(\xi,\eta)$, which is obtained
from eq. \p{aphys1} with the use of the expression \p{sol},
\begin{equation}\label{aphys2}
S=-{i\over 2}\int d^3\xi d^2\eta~{D^\alpha\Phi
D_\alpha\Phi\over{1-{1\over
8}(D^2\Phi)^2}+\sqrt{1+\partial_a\Phi\partial^a\Phi(1-{1\over
16}(D^2\Phi)^2)}}+\int d^3\xi\cdot 1\, .
\end{equation}
One can easily check that in the bosonic limit, when the fermionic
$D_\alpha\Phi|_{\eta=0}$ and auxiliary field
$D^2\Phi|_{\eta=0}$ degrees of freedom are zero, the action
\p{aphys2} reduces to the gauge--fixed Nambu--Goto action
for a membrane in $D=4$
$$
S=\int d^3\xi\sqrt{1+\partial_a\phi\partial^a\phi},
$$
where $\phi(\xi)=\Phi(\xi,\eta)|_{\eta=0}$.

\subsection{Supersymmetry properties of the $d=3$ field theory}
As in the superparticle case of Section 2, the physical gauge
conditions \p{static2}, and therefore the action \p{aphys}, are
invariant under the following combination of the
target--superspace global supersymmetry transformations
\p{SUSY} and the worldvolume superdiffeomorphisms
\begin{equation}\label{wsusyg1}
\delta\eta^\alpha=\epsilon^{1\alpha}, \quad
\delta\xi^m=i\bar\eta\gamma^m\epsilon^1+i\bar\Psi(\xi,\eta)\gamma^m\epsilon^2,
\end{equation}
where
$\epsilon^{\underline\alpha}=(\epsilon^{1\alpha},\epsilon^2_\alpha)$
are two constant parameters of target--space supersymmetry
\p{SUSY} which is seen by the superworldvolume observer as $N=2$,
$d=3$ supersymmetry.

The superfields $\Psi_\alpha(\xi,\eta)$ and $\Phi(\xi,\eta)$
\p{ss}, \p{X3} transform under \p{wsusyg1} in the following way
\begin{equation}\label{Psi}
\delta\Psi_\alpha=-\epsilon^{1\beta}D_\beta\Psi_\alpha
+\epsilon_\alpha^2+i\bar\epsilon^2\gamma^m\Psi\partial_m\Psi_\alpha,
\end{equation}
\begin{equation}\label{phi}
\delta\Phi=-\epsilon^{1\alpha}D_\alpha\Phi+
2i\eta^\alpha\epsilon^2_\alpha+i\bar\epsilon^2\gamma^m\Psi\partial_m\Phi.
\end{equation}
We see that under the
$\epsilon^2$--supersymmetry transformations
$\Psi_\alpha(\xi,\eta)$ and $\Phi(\xi,\eta)$ indeed
vary in a nonlinear manner as Goldstone fields, while under
$\epsilon^1$--supersymmetry they transform as ordinary scalar
superfields. Hence, the $N=2$, $d=3$ supersymmetry of the
superfield action \p{aphys1} is spontaneously broken down to
$N=1$.

\subsection{Relationship to the Goldstone superfield action of \cite{ik1}}

The form of the gauge--fixed supermembrane action \p{aphys1},
\p{aphys2} differs from the Goldstone superfield action
constructed in
\cite{ik1}, because the fields involved in the construction of
these actions are different. The Goldstone superfields of the
former action correspond to the nonlinear realization of
spontaneously broken supersymmetry, while the latter is
constructed with a Goldstone superfield of a `linear' realization.

The Goldstone superfields of the two realizations are related as
follows \cite{ik1}
\begin{equation}\label{psizeta}
\Psi_\alpha={\zeta_\alpha\over{ 1+D^2{\cal F}}}, \quad
\zeta_\alpha=D_\alpha \rho(\xi,\eta),
\end{equation}
where $\rho(\xi,\eta)$ is a scalar superfield and
\begin{equation}\label{tilphi}
{\cal F}={1\over 2} {\zeta^2\over{1+\sqrt{1+D^2\zeta^2}}}.
\end{equation}
From \p{psizeta} and \p{tilphi} we get
\begin{equation}\label{psi2}
\Psi^2=4{\zeta^2\over{(1+\sqrt{1+D^2\zeta^2})^2}}, \quad
D^2\Psi^2=4{D^2\zeta^2\over{(1+\sqrt{1+D^2\zeta^2})^2}}+ \cdots,
\end{equation}
where dots stand for irrelevant terms proportional to
$\zeta$ and $\zeta^2$.

Substituting \p{psi2} into \p{aphys1} we arrive at the action of
\cite{ik1}
$$
S=2i\int d^3\xi d^2\eta {\zeta^2\over{1+\sqrt{1+D^2\zeta^2}}}+\int
d^3\xi \cdot 1\,.
$$
This concludes the relationship of the superembedding description
with other formulations of supermembrane dynamics.

\section{Conclusion}
In this paper, with the examples of the massive $N=2$, $D=2$
superparticle and the $N=1$, $D=4$ supermembrane, we have
demonstrated how starting from the superembedding formulation of
superbrane dynamics one arrives, upon gauge fixing worldvolume
superdiffeomorphisms, at an effective nonlinear field theory on
the brane superworldvolume with partially broken global
supersymmetry. The latter is non--linearly realized on the
superfields composed of supermultiplets transforming linearly
under unbroken supersymmetry.

When the superembedding condition does not put the superbrane
theory on the mass shell there is a generic prescription of how
to construct superbrane actions of the form \p{a1} in the
superembedding approach
\cite{dghs92,pt,hrs}  using a corresponding
Wess--Zumino--like form, which is closed modulo the
superembedding condition \p{emb}. In the case of the
$N=1$, $D=4$ supermembrane we have shown how by consecutive
gauge fixing local worldvolume symmetries, eliminating auxiliary
worldvolume superfields and solving for the superembedding
condition one reduces this generic covariant superfield action to
the $N=1$, $d=3$ Goldstone superfield action exhibiting the
mechanism of partial breaking of $N=2$ global supersymmetry. By
construction this action is directly related to the conventional
supermembrane component action in the physical gauge \cite{achu},
the physical degrees of freedom forming a Goldstone scalar
supermultiplet. We have also demonstrated how the superembedding
action is related to the Goldstone superfield action of
\cite{ik1}.

The superembedding approach thus provides us with a systematic way
of deriving superfield actions for Goldstone superfields of the
method of nonlinear realizations, which have so far been unknown,
and establishes their direct link to the superbrane actions. As
we have mentioned in the Introduction, the actions with partial
supersymmetry breaking have been constructed for a different type
of Goldstone superfields which appear in the method of `linear'
realizations \cite{bg2}--\cite{ketov}. The relation of fermionic
sectors of these actions with the fermionic sectors of
corresponding gauge fixed superbrane actions in general still
remains an open problem.

It should be possible to extend the methods of this paper to more
physically interesting cases, in particular, to the construction
of the covariant superembedding action for a space--filling
Dirichlet 3--brane in an $N=2$, $D=4$ target superspace, whose
dynamics is also described by the off--shell superembedding
condition. A gauge fixed version of this action should be an
action for a $D=4$ supersymmetric Dirac--Born--Infeld field theory
with partially broken $N=2$ supersymmetry described in terms of
Goldstone superfields of the method of nonlinear realizations
\cite{bg2}.

The methods of superembedding and nonlinear realizations are also
applicable to the description of superbranes in
AdS--superbackgrounds and to the derivation of actions for
effective field theories on the AdS boundary whose simple form is
still lacking.

Another direction of research can be connected with studying
partial breaking of {\it local} supersymmetry in supergravity
theories. The embedding of curved supersurfaces into curved
supergravity backgrounds governed by the superembedding condition
\p{emb} seems to be a natural basis for studying mechanisms of
local supersymmetry breaking, which can also be related to the
problem of finding supersymmetric versions of brane world
scenarii.

\bigskip
\noindent
{\bf Acknowledgements.} The authors would like to thank J. Bagger,
I. Bandos, S. Krivonos, S. Kuzenko, E. Ivanov and B. Zupnik for
stimulating discussions. This work was partially supported by the
European Commission TMR Programme ERBFMPX-CT96-0045 to which the
authors are associated. D.S. is grateful to the Organizers of the
Program ``Duality, String Theory and M-Theory" at the Erwin
Schr\"edinger International Institute for Mathematical Physics
(Vienna) for hospitality extended to him during the final stage
of this work.

\renewcommand\theequation{A.\arabic{equation}}
\appendix
\section*{Appendix: The $N=2$, $D=2$ massive superparticle in the
light--cone gauge}

Below we demonstrate that when the $N=2$, $D=2$ massive
superparticle action is chosen in the form \p{2n2d}, the
worldline local superreparametrization \p{wsusy} allows one to
impose in this action the light--cone gauge condition instead of
the static gauge condition \p{static1}.

Remember that the action \p{2n2d} is obtained from a more general
action \p{d2} upon gauge fixing one of the independent
superdiffeomorphisms \p{dif} of the latter by putting the
worldline supervielbein component $f(\tau,\eta)$ equal to the
Grassmann--odd coordinate $\eta$ ($f(\tau,\eta)=\eta)$. This
reduces the superdiffeomorphism group \p{dif} down to \p{wsusy}.
To gauge fix the latter we impose the light--cone gauge condition
\begin{equation}\label{lc}
X^-=(X^0-X^1)=\tau.
\end{equation}
Then, in view of the Majorana form of the Gamma--matrices
\p{Gamma1}, the superembedding condition \p{emb1} reduces to the
following two equations
\begin{equation}\label{-}
DX^-=i\eta=iD(\Theta^1-\Theta^2)(\Theta^1-\Theta^2),
\end{equation}
\begin{equation}\label{+}
DX^+\equiv D(X^0+X^1)=iD(\Theta^1+\Theta^2)(\Theta^1+\Theta^2),
\end{equation}

Solving for eq. \p{-} we get the light--cone gauge condition for
the Grassmann--odd coordinates
\begin{equation}\label{lcg}
\Theta^1-\Theta^2\equiv\Theta^-=\pm\eta.
\end{equation}
The equation \p{+} can be solved for
$\Psi^+\equiv\Theta^1+\Theta^2$ using the Bagger--Galperin trick
\cite{bg2}. From \p{lcg} we get
 \begin{equation}\label{step1}
\Psi^+=-i{{DX^+}\over{D\Psi^+}}.
 \end{equation}
Now take the $D$--derivative of \p{step1}
\begin{equation}\label{step2}
D\Psi^+=-{{\partial_\tau
X^+}\over{D\Psi^+}}+{{DX^+\partial_\tau\Psi^+}\over{(D\Psi^+})^2}.
\end{equation}
Examining the eq. \p{step2} we find that because $DX^+$ is
Grassmann--odd $((DX^+)^2\equiv 0)$ the second term of the right
hand side of \p{step2} does not contribute to the right hand side
of \p{step1} and to the first term of \p{step2}, when we
substitute $D\Psi^+$ in these terms with its recursive relation
\p{step2}. This allows us to write down ``effective" relations
\begin{equation}\label{step3}
(D\Psi^+)_{eff}=-{{\partial_\tau X^+}\over{(D\Psi^+)_{eff}}},
\quad
 \Psi^+=-i{{DX^+}\over{(D\Psi^+)_{eff}}}.
\end{equation}
From the first equation in \p{step3} we get (up to an irrelevant
sign) $(D\Psi^+)_{eff}=\sqrt{\partial_\tau X^+}$, then the second
equation takes the form
\begin{equation}\label{step4}
\Psi^+=-i{{DX^+}\over{\sqrt{\partial_\tau X^+}}},
\end{equation}
which expresses the superfield $\Psi^+$ in terms of $X^+$.

We have thus explicitly solved the superembedding constraints
\p{-}, \p{+} in the light--cone gauge \p{lc}, \p{lcg}. As a result
(up to a total derivative) the superfield action \p{2n2d} reduces
to
\begin{equation}\label{lca}
S=m\int d\tau d\eta ~\Psi^+=-im\int d\tau
d\eta{{DX^+}\over{\sqrt{\partial_\tau X^+}}}.
\end{equation}
One can easily verify that eq. \p{lca} is the $N=1$ worldline
superfield form of the component action which one obtains by
imposing the light--cone gauge in the standard component action
for the $N=2$, $D=2$ superparticle
\begin{equation}\label{com}
S=m\int d\tau\left[\sqrt{(\partial_\tau x^{\underline
a}-i\partial_\tau\bar\theta\Gamma^{\underline
a}\theta)(\partial_\tau x_{\underline
a}-i\partial_\tau\bar\theta\Gamma_{\underline
a}\theta)}-i\partial_\tau\bar\theta\Gamma^2\theta \right],
\end{equation}
or in the first--order form
\begin{equation}\label{com1}
S=\int d\tau ~\left[p_{\underline a}(\partial_\tau x^{\underline
a}-i\partial_\tau\bar\theta\Gamma^{\underline
a}\theta)-{e(\tau)\over 2}(p_{\underline a}p^{\underline
a}-m^2)-i\partial_\tau\bar\theta\Gamma^2\theta \right]
\end{equation}

The superfield $X^+(\tau,\eta)$ is composed from the light--cone
coordinates $x^+(\tau)$ and $\theta^+(\tau)$ of \p{com} as follows
 $$
X^+(\tau,\eta)=x^+(\tau)+i\eta\theta^+\sqrt{\partial_\tau x^+}.
 $$

\end{document}